\documentclass{aa}  
\usepackage{natbib,twoopt}
\usepackage[breaklinks=true]{hyperref}
\bibpunct{(}{)}{;}{a}{}{,}
\makeatletter
\newcommandtwoopt{\citeads}[3][][]{\href{http://adsabs.harvard.edu/abs/#3}
{\def\hyper@linkstart##1##2{}
\let\hyper@linkend\@empty\citealp[#1][#2]{#3}}}
\newcommandtwoopt{\citepads}[3][][]{\href{http://adsabs.harvard.edu/abs/#3}
{\def\hyper@linkstart##1##2{}
\let\hyper@linkend\@empty\citep[#1][#2]{#3}}}
\newcommandtwoopt{\citetads}[3][][]{\href{http://adsabs.harvard.edu/abs/#3}
{\def\hyper@linkstart##1##2{}
\let\hyper@linkend\@empty\citet[#1][#2]{#3}}}
\newcommandtwoopt{\citeyearads}[3][][]
{\href{http://adsabs.harvard.edu/abs/#3}
{\def\hyper@linkstart##1##2{}
\let\hyper@linkend\@empty\citeyear[#1][#2]{#3}}}
\makeatother
\usepackage{graphicx}
\usepackage{txfonts}

\begin{document}

   \title{``Zombie'' or active? An alternative explanation to the properties of star-forming galaxies at high redshift}

   \author{F. G. Saturni\inst{\ref{inst1},\ref{inst2}}
          \and
          M. Mancini\inst{\ref{inst3}}
          \and
          E. Pezzulli\inst{\ref{inst1},\ref{inst4},\ref{inst5}}
          \and
          F. Tombesi\inst{\ref{inst1},\ref{inst6},\ref{inst7},\ref{inst8}}
          }

   \institute{INAF -- Osservatorio Astronomico di Roma, Via Frascati 33, 00040 Monte Porzio Catone (RM), Italy.\\
              \email{francesco.saturni@inaf.it}\label{inst1}
         \and
             Space Science Data Center, Agenzia Spaziale Italiana, Via del Politecnico snc, 00133 Roma, Italy.\label{inst2}
             \and
             ASTRON -- Netherlands Institute for Radio Astronomy, Oude Hoogeveensedijk 4, 7991 PD Dwingeloo, The Netherlands.\label{inst3}
             \and
             Dipartimento di Fisica, Universit{\`a} degli Studi di Roma ``La Sapienza'', P.le A. Moro 5, 00185 Roma, Italy.\label{inst4}
             \and
             INFN, Sezione di Roma I, P.le A. Moro 2, 00185 Roma, Italy.\label{inst5}
             \and
             Dipartimento di Fisica, Universit{\`a} di Roma ``Tor Vergata'', Via della Ricerca Scientifica 1, 00133 Roma, Italy.\label{inst6}
             \and
             Department of Astronomy, University of Maryland, College Park, MD 20742, USA.\label{inst7}
             \and
             NASA/Goddard Space Flight Center, Code 662, Greenbelt, MD 20771, USA.\label{inst8}
             }

   \date{Received 2018 Apr 19; accepted 2018 Jun 18}

  \abstract  
   {Star-forming galaxies at high redshift show anomalous values of infrared excess, which can be described only by extremizing the existing relations between the shape of their ultraviolet continuum emission and their infrared-to-ultraviolet luminosity ratio, or by constructing {ad hoc} models of star formation and dust distribution.}
   {We present an alternative explanation, based on unveiled AGN activity, of the existence of such galaxies. The scenario of a weak AGN lends itself naturally to explain the observed spectral properties of these high-z objects in terms of a continuum slope distribution and not altered infrared excesses.}
   {To this end, we directly compare the infrared-to-ultraviolet properties of high-redshift galaxies to those of known categories of AGN (quasars and Seyferts). We also infer the characteristics of their possible X-ray emission.}
   {We find a strong similarity between the spectral shapes and luminosity ratios of AGN with the corresponding properties of such galaxies. In addition, we derive expected X-ray fluxes that are compatible with the energetics from AGN activity.}
   {We conclude that a moderate AGN contribution to the UV emission of such high-$z$ objects is a valid alternative to explain their spectral properties. Even the presence of an active nucleus in each source would not violate the expected quasar statistics. Furthermore, we suggest that the observed similarities between anomalous star-forming galaxies and quasars may provide a benchmark for future theoretical and observational studies on the galaxy population in the early Universe.}


   \maketitle

\section{Introduction}

The estimate of the galactic dust content and properties in objects at high redshift ($z>5$) is crucial for assessing the details of the star formation history in galaxies. However, it is not always feasible to allow the measurement of the
far-infrared (FIR) flux emission, due to low instrumental sensitivity or lack of spectral coverage  \citep[e.g.,][and references therein]{Bouwens2014,Rogers2014,Finkelstein2015,Finkelstein2015b,Bowler2015,Capak2015,Bouwens2016,Barisic2017}. Under the assumption of radiative equilibrium, it is possible to correlate the FIR emission to the amount of ultraviolet (UV) flux absorbed by dust and reprocessed at longer wavelengths since the dust extinction of reasonable absorption models peaks in the UV band \citep{Weingartner2001,Bianchi2007,draine2011}. Therefore, it is common practice to quantify the extinction at such redshifts using the galaxy-emission properties in the rest-frame UV band \citep{Bouwens2014,Bouwens2015a,Finkelstein2015}.

\begin{table*}
\renewcommand{\arraystretch}{1.3}
\resizebox{\textwidth}{!}{
\begin{tabular}{lccccccc}
\hline
\hline
ID & R.A. (h m s) & dec (d m s) & $z$ & $L_{\rm FIR}$ $(L_\odot)$ & $L_{\rm UV}$ $(L_\odot)$ & $\beta$ & IRX\\
\hline
HZ1 & 09 59 53.25 & +02 07 05.43 & 5.69 & $<$10.32 & $11.210 \pm 0.010$ & $-1.92^{+0.14}_{-0.11}$ & $<-0.036$ \\
HZ2 & 10 02 04.10 & +01 55 44.05 & 5.67 & $<$10.30 & $11.150 \pm 0.010$ & $-1.82 \pm 0.10$ & $<-0.034$ \\
HZ3 & 10 00 09.43 & +02 20 13.86 & 5.55 & $<$10.53 & $11.080 \pm 0.010$ & $-1.72^{+0.12}_{-0.15}$ & $<-0.022$ \\
HZ4 & 09 58 28.52 & +02 03 06.74 & 5.54 & $11.13 \pm 0.54$ & $11.280 \pm 0.010$ & $-2.06^{+0.13}_{-0.15}$ & $-0.006 \pm 0.292$ \\
HZ5 & 10 00 51.60 & +02 34 57.55 & 5.31 & $<$10.30 & $11.450 \pm 0.010$ & $-1.01^{+0.06}_{-0.12}$ & $<-0.046$ \\
HZ5a & 10 00 51.52 & +02 34 59.29 & 5.31 & $<$10.30 & $<$10.370 & --- & --- \\
HZ6 & 10 00 21.50 & +02 35 11.08 & 5.29 & $11.13 \pm 0.23$ & $11.470 \pm 0.100$ & $-1.14^{+0.12}_{-0.14}$ & $-0.013 \pm 0.063$ \\
HZ6a & --- & --- & 5.29 & $10.26 \pm 0.23$ & $11.110 \pm 0.070$ & $-0.59^{+1.05}_{-1.12}$ & $-0.035 \pm 0.058$ \\
HZ6b & --- & --- & 5.29 & $10.87 \pm 0.23$ & $11.000 \pm 0.070$ & $-1.50^{+1.05}_{-1.22}$ & $-0.005 \pm 0.058$ \\
HZ6c & --- & --- & 5.29 & $10.79 \pm 0.23$ & $10.810 \pm 0.070$ & $-1.30^{+0.51}_{-0.37}$ & $-0.001 \pm 0.058$ \\ 
HZ7 & 09 59 30.48 & +02 08 02.81 & 5.25 & $<$10.35 & $11.050 \pm 0.020$ & $-1.39^{+0.15}_{-0.17}$ & $<-0.028$ \\
HZ8 & 10 00 04.06 & +02 37 35.81 & 5.15 & $<$10.26 & $11.040 \pm 0.020$ & $-1.42^{+0.19}_{-0.18}$ & $<-0.032$ \\
HZ8W & 10 00 03.97 & +02 37 36.23 & 5.15 & $<$10.26 & $10.570 \pm 0.040$ & $-0.10 \pm 0.29$ & $-0.013 \pm 0.002$ \\
HZ9 & 09 59 51.70 & +02 22 42.16 & 5.55 & $11.54 \pm 0.19$ & $10.950 \pm 0.020$ & $-1.59^{+0.22}_{-0.23}$ & $0.023 \pm 0.036$ \\ 
HZ10 & 10 00 59.30 & +01 33 19.53 & 5.66 & $11.94 \pm 0.08$ & $11.140 \pm 0.020$ & $-1.92^{+0.24}_{-0.17}$ & $0.030 \pm 0.007$ \\
HZ10W & --- & --- & 5.66 & $11.64 \pm 0.08$ & $10.230 \pm 0.050$ & $-1.47^{+0.77}_{-0.44}$ & $0.056 \pm 0.009$\\
\hline
\end{tabular}
}
\caption{Basic properties of the high-$z$ galaxies discovered by \citet{Capak2015} and reanalyzed by \citet{Barisic2017}. The objects are ordered by ID number. Values preceded by ``$<$'' represent an upper limit in the corresponding quantity, while cells with ``---'' denote missing data from the original publications.}
\label{table:data}
\end{table*}

The recent detection of FIR emission from high-$z$ galaxies \citep[$z>5$;][]{Finkelstein2015,Finkelstein2015b,Bowler2015,Capak2015,Waters2016,Bouwens2016,Knudsen2017} provides better constraints on their dust properties, such as extinction. In particular, the study of galaxies with non-negligible star formation rates (SFRs) shows that their intrinsic wavelength spectral index $\beta_{\rm int}$, i.e.,  the slope of the power-law spectrum used to describe the galactic UV emission, is nearly constant with values between $-2.5$ and $-2.3$ \citep{Meurer1999,Wilkins2013,Bouwens2014,Reddy2015,Forrest2016,Mancini2016,Narayanan2017,Reddy2017,Popping2017,Cullen2017}, although \citet{Reddy2015}, \citet{Casey2014} and \citet{Forrest2016} found a value compatible with $\beta > -2.3$ and \citet{Talia2015} estimated $\beta \sim -3.02$.

\citet{Meurer1999} initially suggested the possibility of correlating the difference between the observed and the expected $\beta$, which is used to describe the galactic UV continuum, to the extinction coefficient at 1600 \AA~(hereafter $A_{1600}$). This procedure is made possible by  the independence of the intrinsic spectrum of star-forming galaxies from the stellar-population age and by its weak correlation with the stellar metallicity \citep[see, e.g.,][and references therein]{Meurer1999,mancini2015}. Furthermore, it is possible to correlate $A_{1600}$ with the infrared excess (IRX) defined as
\begin{equation}\label{eqn:irx}
{\rm IRX} = \log{\left(
\frac{L_{\rm FIR}}{L_{\rm UV}}
\right)},
\end{equation}
where $L_{\rm FIR}$ is the FIR galactic luminosity integrated from 8 $\mu$m to 1000 $\mu$m and $L_{\rm UV}$ is the monochromatic luminosity at 1600 \AA~in the rest frame. The IRX-to-$\beta$ relation obtained in this way only depends on the dust mass (i.e., the normalization of the dust-extinction law), the dust properties (i.e., the shape of such a law), and the intrinsic spectral index $\beta_{\rm int}$ corrected for the dust absorption.

This relation appears to be universal for star-forming galaxies \citep[e.g.,][]{Meurer1999}. However, \citet{Capak2015} observed a population of high-$z$ star-forming galaxies that appear to strongly deviate from the usual relations, showing lower IRX with respect to that expected on the basis of their observed spectral slope. These sources were selected  to measure [C {\scriptsize II}] gas and dust emission in Lyman-break galaxies (LBGs) detected in the  2 deg$^2$ Cosmological Evolution Survey \citep[COSMOS;][]{Scoville2007} in order to explore the average properties of the interstellar medium (ISM) at high redshift. The reanalysis of the \citet{Capak2015} data done by \citet{Barisic2017} partially relaxed the tension between their IRXs and spectral slopes, but without eliminating it, especially for the case of upper limits to the IRX. These objects populate a region of the IRX--$\beta$ plane that is typical of quiescent galaxies with  negligible or no star formation (``dead'') despite having SFRs from tens to hundreds of $M_{\odot}$ yr$^{-1}$ as measured by their [C {\scriptsize II}] emission \citep{Capak2015}, and could therefore be denoted as ``zombie'' galaxies.

Several scenarios have been proposed to explain how star-forming galaxies can deviate from the usual IRX-to-$\beta$ relations (\citealt{Bouwens2016,Ferrara2017}; see also \citealt{Ouchi1999}). However, the observed discrepancy with the predicted IRXs would either imply a strong bias in the determination of the dust temperature \citep{Bouwens2016} or extreme physical conditions in the dust clouds \citep{Ferrara2017}. In this work, we formulate the alternative hypothesis that the high-$z$ galaxies with anomalous IRX-to-$\beta$ discovered by \citet{Capak2015} are actually hosting AGN activity. This scenario allows us to consider objects in which the UV flux is dominated by accretion-disk processes \citep[e.g.,][]{Salpeter1964,Zeldovich1965,Shakura1973} and in which the intrinsic UV spectral slope depends on the properties of the central engine (accretion rate, black hole mass, black hole spin) and is therefore not limited to steep values \citep[e.g.,][]{Shankar2016}. In this way, the \citet{Capak2015} objects are no longer constrained  to have anomalous dust properties, but are instead located in the region of the IRX--$\beta$ plane corresponding to their AGN UV continuum.

\begin{figure*}
\centering
\begin{minipage}{.49\textwidth}
\includegraphics[width=\textwidth]{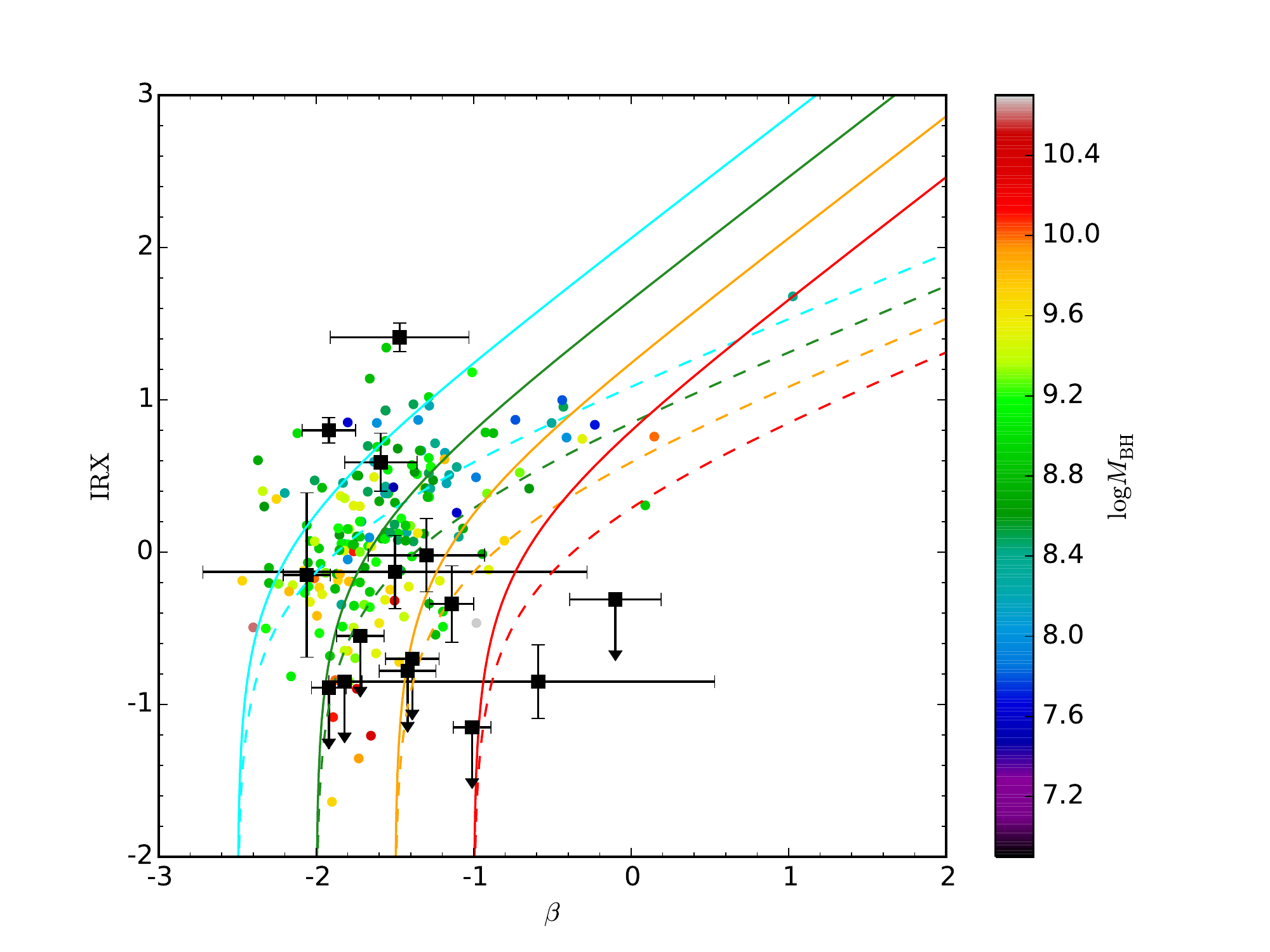}
\end{minipage}
\begin{minipage}{.49\textwidth}
\includegraphics[width=\textwidth]{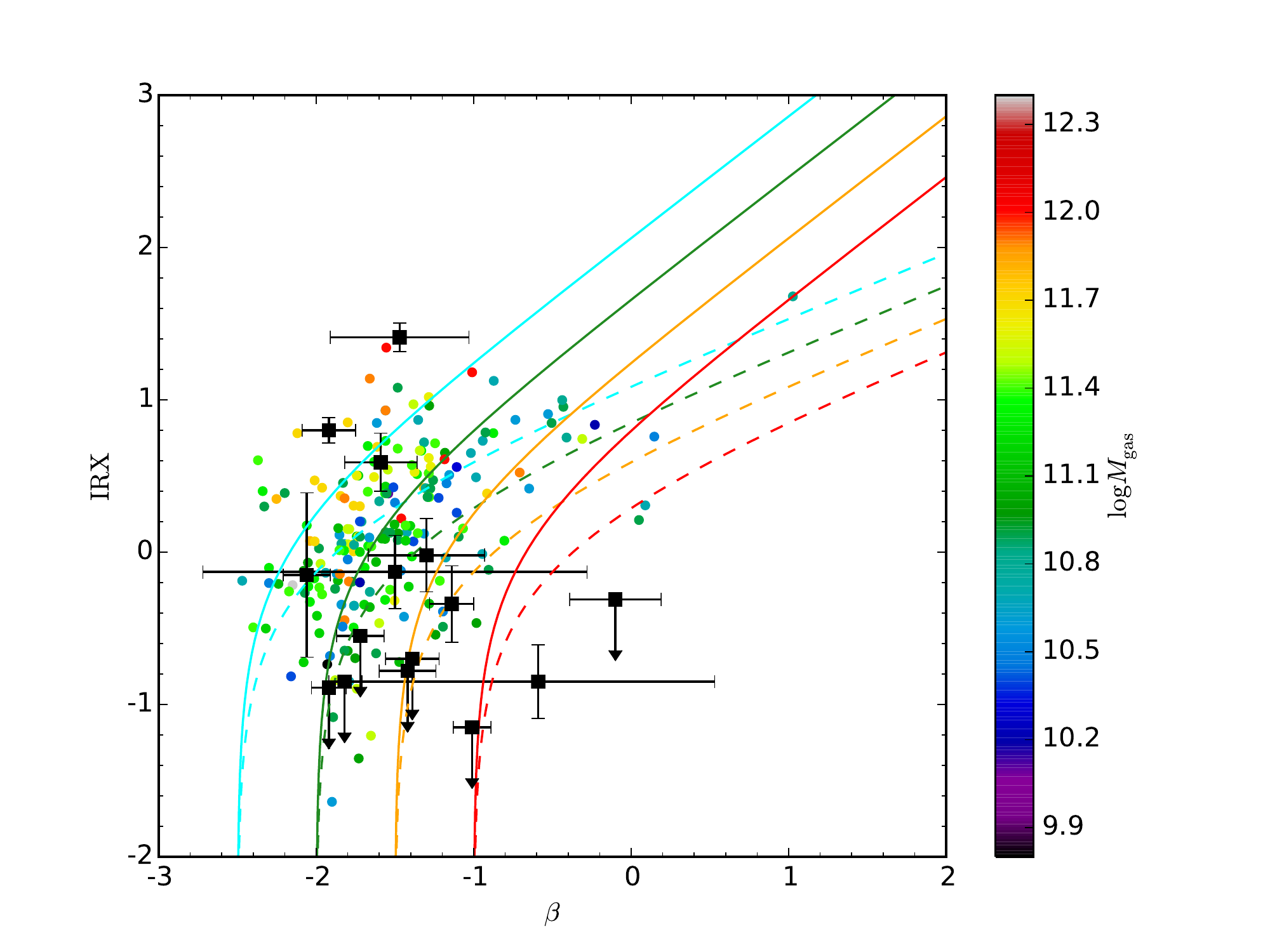}
\end{minipage}
\caption{IRX--$\beta$ plane for quasars.  The data corresponding to the high-$z$ galaxies by \citet{Capak2015} reanalyzed by \citet{Barisic2017} are shown ({\itshape black squares}), along with their 1$\sigma$ errors, superimposed on the HerS/SDSS quasars ({\itshape filled dots}). The empirical IRX-to-$\beta$ relations by \citet[][{\itshape dashed lines}\ignorespaces ]{Pettini1998} and \citet[][{\itshape solid lines}\ignorespaces ]{Meurer1999} computed for values of $\beta_{\rm int}$ from $-2.5$ to $-1$ are also reported. {\itshape Left panel:} HerS/SDSS quasars segregated in logarithmic $M_{\rm BH}$. {\itshape Right panel:} HerS/SDSS quasars segregated in logarithmic $M_{\rm gas}$. Only HerS/SDSS objects with measured $M_{\rm BH}$ and $M_{\rm gas}$ are used.}
\label{fig:irxcolored}
\end{figure*}

The paper is organized as follows.   In Sect. \ref{sec:irxagn} we exploit the idea developed by
\citet{Capak2015} that AGN activity is responsible for the altered IRX-to-$\beta$ of the high-$z$ galaxies. We present a tentative statistical classification of these objects based on their spectral properties in Sect. \ref{sec:cmragn}. We also compute the expected X-ray emission from AGN with properties similar to these sources in Sect. \ref{sec:xraylum}, and compare the fluxes estimated in this way with the current X-ray upper limits of these sources derived from {\itshape Chandra} observations of the COSMOS field. Finally, we summarize and discuss our findings in Sect. \ref{sec:disc}. Throughout the text, we use  ``AGN'' to indicate both low- and high-luminosity objects, and adopt a concordance cosmology with $H_0 = 70$ km s$^{-1}$ Mpc$^{-1}$, $\Omega_{\rm M} = 0.3$, and $\Omega_\Lambda = 0.7$.

\section{The IRX-to-$\beta$ relation for AGN}\label{sec:irxagn}

The sample of 16 high-$z$ galaxies discovered by \citet{Capak2015} is presented in Table \ref{table:data}. It comprises six single sources (HZ1, HZ2, HZ3, HZ4, HZ7, and HZ9), and four multiple systems (HZ5, HZ6, HZ8, and HZ10) that are probably composed of interacting objects. Their redshifts were measured through absorption-line spectroscopy with the instrument DEIMOS at the {\itshape Keck} Observatory. For all but one of them (HZ5a), the measurements or at the least upper limits of their UV spectral slope and IRX are given in \citet{Barisic2017}. In the following, we adopt this data set for our analysis of such objects in search of AGN behavior.

In order to validate the hypothesis that star-forming galaxies at high redshift with anomalous IRX-to-$\beta$ values are hosts of AGN activity, we first define a comparison sample of AGN that has measurements of UV to optical spectral slopes $\beta$, IR luminosities $L_{\rm IR}$, and UV luminosities $L_{\rm UV}$. We find that \citet{Dong2016} constructed such a catalog for quasars, i.e., AGN with $L_{\rm bol} \gtrsim 10^{45}$ erg s$^{-1}$. A similar catalog has been constructed by \citet{Garcia2016} for nearby ($D_{\rm L} < 70$ Mpc) Seyfert galaxies ($L_{\rm bol} \lesssim 10^{45}$ erg s$^{-1}$). By comparing both catalogs  to the \citet{Capak2015} galaxies, we are able to span five orders of magnitude in bolometric luminosity from $10^{43}$ to $10^{48}$ erg s$^{-1}$, thus covering a wide range of AGN energetics.

We note that the selection criteria used in these AGN catalogs significantly differ from those adopted by \citet{Capak2015} and \citet{Barisic2017}, due to the diverse purposes for which the samples were constructed (the study of the link between AGN activity and host-galaxy star formation in the case of \citealt{Dong2016}, and the disentanglement of the FIR emission due to AGN activity from that due to star formation in \citealt{Garcia2016}). Nevertheless, a comparison between AGN and the \citet{Capak2015} objects based on the average properties of such samples is still feasible because we  focus on spectral properties that are nearly unhindered by the nature of the ISM gas. For instance, the zombie galaxy HZ5, though satisfying the object selection performed by \citet{Capak2015}, is actually a weak Type 1 quasar \citep{Brightman2014,Kalfountzou2014}. Therefore, we expect that the different selection criteria among the three samples do not greatly affect our findings.

\subsection{Construction of a quasar comparison sample}

The quasar catalog by \citet{Dong2016} is composed of objects from the Sloan Digital Sky Survey Data Release 7 (SDSS-DR7; \citealt{Abazajian2009,Schneider2010,Shen2011}) and Data Release 10 (SDSS-DR10; \citealt{Ahn2014,Paris2014}) that were observed in the FIR domain by the Spectral and Photometric Imaging Receiver (SPIRE) on board  the {\itshape Herschel Space Observatory} in the framework of the {\itshape Herschel} Stripe 82 Survey (HerS; \citealt{Viero2014}). This survey covers $\sim$29\% (79 deg$^2$) of the SDSS Stripe 82 (270 deg$^2$ on the celestial equator in the southern Galactic cap; \citealt{Adelman2007}) with FIR observations at 250, 350, and 500 $\mu$m.

The joint HerS/SDSS sample includes 207 quasars with the following characteristics:
\begin{itemize}
        \item UV to optical {\itshape ugriz} magnitudes from the SDSS;
        \item near-IR (NIR) $JHK_{\rm s}$ magnitudes from the 2MASS \citep{Skrutskie2006} and UKIDSS \citep{Lawrence2007} surveys;
        \item mid-IR (MIR) fluxes from the AllWISE catalog \citep{Wright2010,Mainzer2011} ranging from $3.4$ to 22 $\mu$m;
        \item {\itshape Herschel} FIR flux measurements in the passbands centered at the wavelengths mentioned above.
\end{itemize}

The spectral energy distribution (SED) of each object has been decomposed in several spectral components, each accounting for the emission from a different source: a power law for the quasar accretion disk, two near- to mid-IR bumps peaking at around $2 - 4$ $\mu$m and $10 - 20$ $\mu$m for the radiation reprocessed by the dusty torus, two galaxy templates (S$b$ for the young stellar component and E for the old one) for the host-galaxy contribution, a Small Magellanic Cloud (SMC) reddening law and a ``graybody'' radiation with power-law emissivity for the FIR excess. In this way, \citet{Dong2016} measured IR and UV luminosities for each quasar in the sample, along with the corresponding frequency spectral slope $\alpha$ which is related to the wavelength slope $\beta$ by
\begin{equation}\label{eqn:alphabeta}
\beta = -(2+\alpha)
\end{equation}
Thanks to the SED analysis, the HerS/SDSS catalog also provides information on several other quasar and host-galaxy  parameters, such as the virial black hole (BH) mass $M_{\rm BH}$ derived from single-epoch relations \citep{Vestergaard2006,Vestergaard2009,Shen2010} and the mass of cold gas $M_{\rm gas}$ stored in the AGN surrounding environment. The former is measured by adopting the quasar continuum luminosities computed from the power-law emission, while the latter is estimated from the long-wavelength tail of the dust emission \citep[see][for details]{Scoville2014}.

Since the SED decomposition in multiple components allows us  to measure quasar intrinsic reddening, this may be used to correct the quantities derived from the fitting procedure. Therefore, $L_{\rm UV}$ and $\beta$ given in the HerS/SDSS sample are {\itshape intrinsic} to the quasars, whereas the corresponding {\itshape observed} quantities (i.e., with no correction for internal reddening applied) are needed in the construction of the IRX-to-$\beta$ relation. In order to recover the observed $L_{\rm UV}$ and $\beta$, we apply backward the de-reddening procedure adopted by \citet{Dong2016}. We individually de-correct both $L_{\rm UV}$ and $\beta$ through a SMC extinction template \citep{Pei1992} scaled to the value of intrinsic reddening $A_{\rm int}$ of each quasar, and use these reddened quantities in the evaluation of the quasar IRX-to-$\beta$ relation.

The resulting sample of quasar IRX and spectral slopes is shown in Fig. \ref{fig:irxcolored}. At  first glance, it is evident as all but three high-$z$ galaxies by \citet{Capak2015} fall inside the region of the IRX--$\beta$ diagram populated by quasars. For comparison, we compute the IRX-to-$\beta$ relations by \citet{Pettini1998} and \citet{Meurer1999} for several values of the intrinsic spectral slope $\beta_{\rm int}$ from $-2.5$ to $-1$, and show them superimposed on the observed IRX-to-$\beta$ values for high-$z$ galaxies and quasars.

It should  be noted that the HerS/SDSS quasars tend to segregate in the IRX--$\beta$ plane according to both their $M_{\rm BH}$ (left panel of Fig. \ref{fig:irxcolored}) and $M_{\rm gas}$ (right panel of Fig. \ref{fig:irxcolored}). In particular, quasars hosting more massive black holes have lower IRXs, whereas objects with higher gas masses show steeper intrinsic slopes coupled to slightly higher IRXs. It is unclear whether the tendency for quasars with higher $M_{\rm BH}$ to have lower IRXs is  due to a physical mechanism such as quasar feedback \citep[e.g.,][]{Cattaneo2009,Fabian2012}, since the main driver of feedback appears to be the quasar Eddington ratio $L_{\rm bol}/L_{\rm edd}$ rather than the black hole mass (\citealt{Ricci2017a}; see also \citealt{Ishibashi2016,Ishibashi2017}). Such a segregation could  equally be produced by a selection bias in the observation of quasars with massive black holes, for example by preferentially finding objects with bluer spectra at the high end of the $M_{\rm BH}$ distribution due to their lower Eddington ratio.

\begin{figure}
\includegraphics[width=.75\columnwidth,angle=-90]{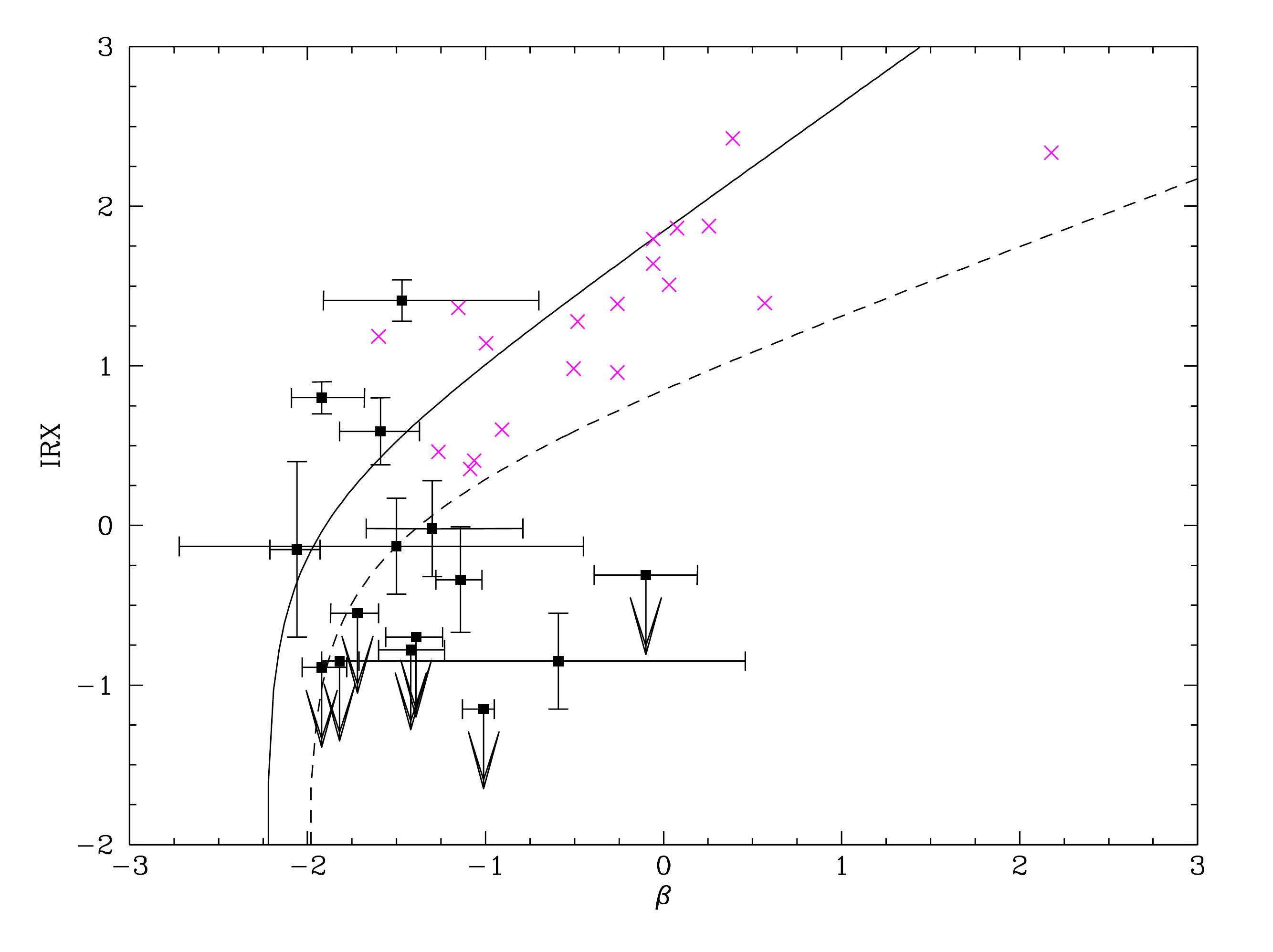}
\caption{IRX--$\beta$ plane for Seyfert galaxies. The data by \citet{Barisic2017} are shown ({\itshape black squares}), along with their 1$\sigma$ errors, superimposed on the sample of Seyfert galaxies by \citet[][{\itshape magenta crosses}\ignorespaces ]{Garcia2016}. For comparison, the theoretical IRX-to-$\beta$ relation by \citet{Meurer1999} with $\beta_{\rm int} = -2.23$ ({\itshape solid line}) and the relation by \citet{Pettini1998} with $\beta_{\rm int} = -2$ ({\itshape dashed line}) are also reported.}
\label{fig:IRXSy}
\end{figure}

Conversely, a high $M_{\rm gas}$ may be associated with gas-rich quasar surrounding environments, resulting in high-luminosity objects which tend to have bluer spectra and hence steeper $\beta_{\rm int}$ \citep[e.g.,][]{Shankar2016}. Since high gas masses also enhance the star formation and dust content of the host galaxy \citep[e.g.,][]{Kennicutt2012}, the tendency of quasars with higher $M_{\rm gas}$ to segregate towards steeper UV spectral slopes and higher IRXs is explained by the simultaneous presence of a higher AGN and stellar luminosity, and a higher dust mass. We note that future observations of high-$z$ galaxies could benefit from using these tendencies as diagnostics  to infer the presence of a centrally located AGN powering part of the UV emission.

\subsection{Construction of a Seyfert comparison sample}

The Seyfert comparison sample has been constructed in \citet{Garcia2016} by selecting objects from the Revised Shapley -- Ames (RSA) catalog \citep{Sandage1987} that were identified as Seyfert galaxies by \citet{Maiolino1995}. These objects were selected in order to have {\itshape Herschel}/Photometric Array Camera (PACS) imaging in at least two bands and SPIRE photometric observations. The final catalog by \citet{Garcia2016} comprises 33 galaxies (15 Sy 1 and 18 Sy 2), of which we only use those with near- and far-UV (NUV and FUV) magnitudes available in the literature in order to compute UV spectral slopes and luminosities. We therefore select Seyfert galaxies from this catalog by performing a search on the NASA Extragalactic Database\footnote{Available at {\ttfamily ned.ipac.caltech.edu}.} (NED) for objects with both NUV and FUV measurements.

Our final sample is composed of 19 Seyferts with
\begin{itemize}
        \item NUV and FUV AB magnitudes measured by the {\itshape Galaxy Evolution Explorer} ({\itshape GALEX}) satellite for UV surveys \citep{Martin2005,Gildepaz2007};
        \item MIR spectroscopy with high angular resolution \citep[see][and refs. therein]{Garcia2016};
        \item {\itshape Herschel} FIR photometric measurements from 70 $\mu$m to 500 $\mu$m.
\end{itemize}
All these objects have IR SEDs fitted by \citet{Garcia2016} with graybody components which allow us to compute $L_{\rm FIR}$. Coupling these luminosities to the UV magnitudes, we are able to compute the IRX for this subsample, which we show in Fig. \ref{fig:IRXSy} as a function of the corresponding $\beta$ compared to the high-$z$ galaxies by \citet{Capak2015}. A visual inspection reveals that these Seyferts have shallower UV spectral slopes and higher IRXs with respect to quasars, thus populating the region at $-1 \lesssim \beta \lesssim 0$ and ${\rm IRX} \gtrsim 0$ which contains neither high-$z$ galaxies nor HerS/SDSS quasars (see Fig. \ref{fig:irxcolored}). This suggests that a possible AGN activity in the \citet{Capak2015} objects must have properties closer to efficient high-luminosity AGN than to weak ones.

\begin{figure*}
\begin{minipage}{.49\textwidth}
\includegraphics[width=.75\textwidth,angle=-90]{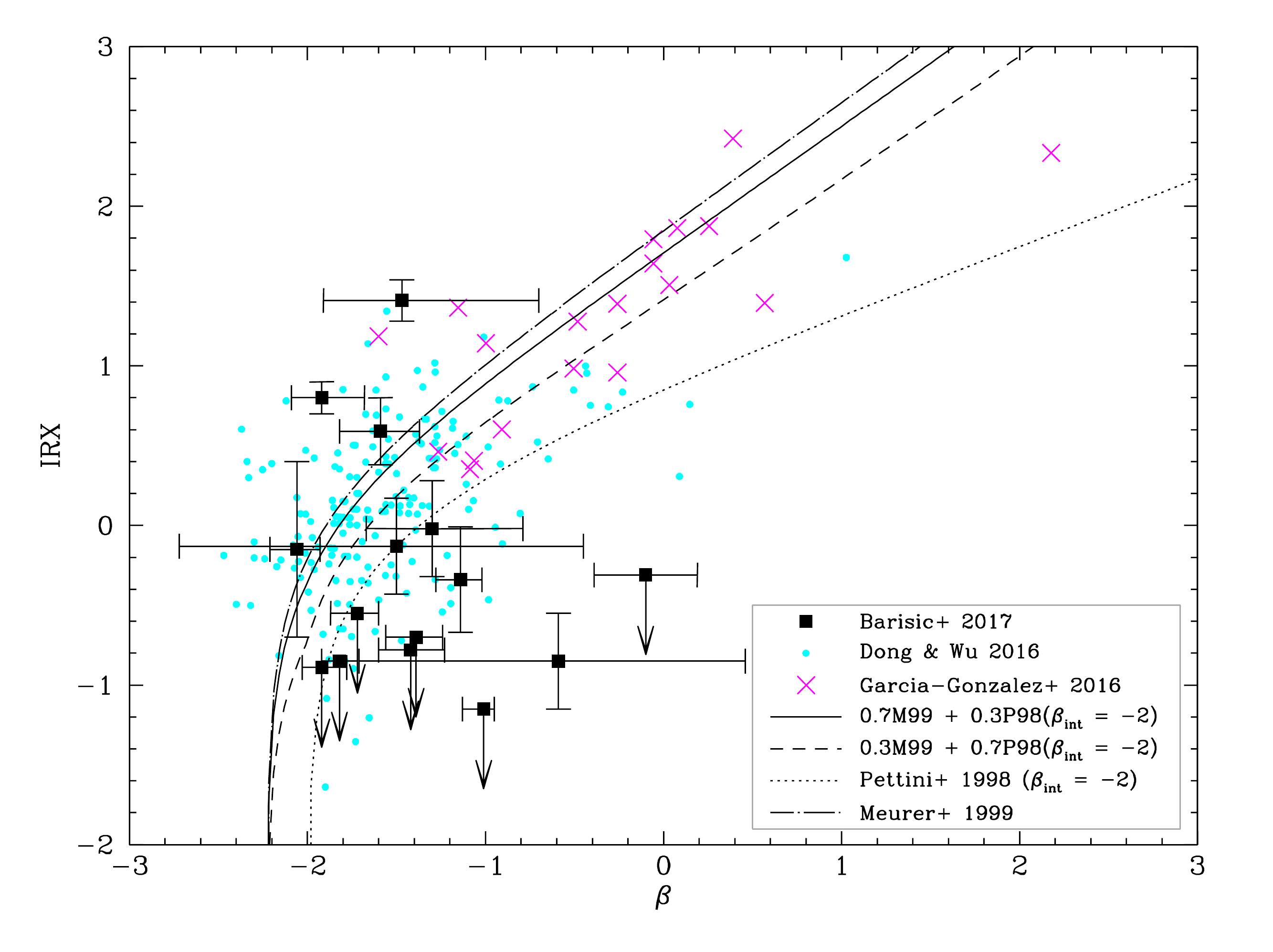}
\end{minipage}
\begin{minipage}{.49\textwidth}
\includegraphics[width=.75\textwidth,angle=-90]{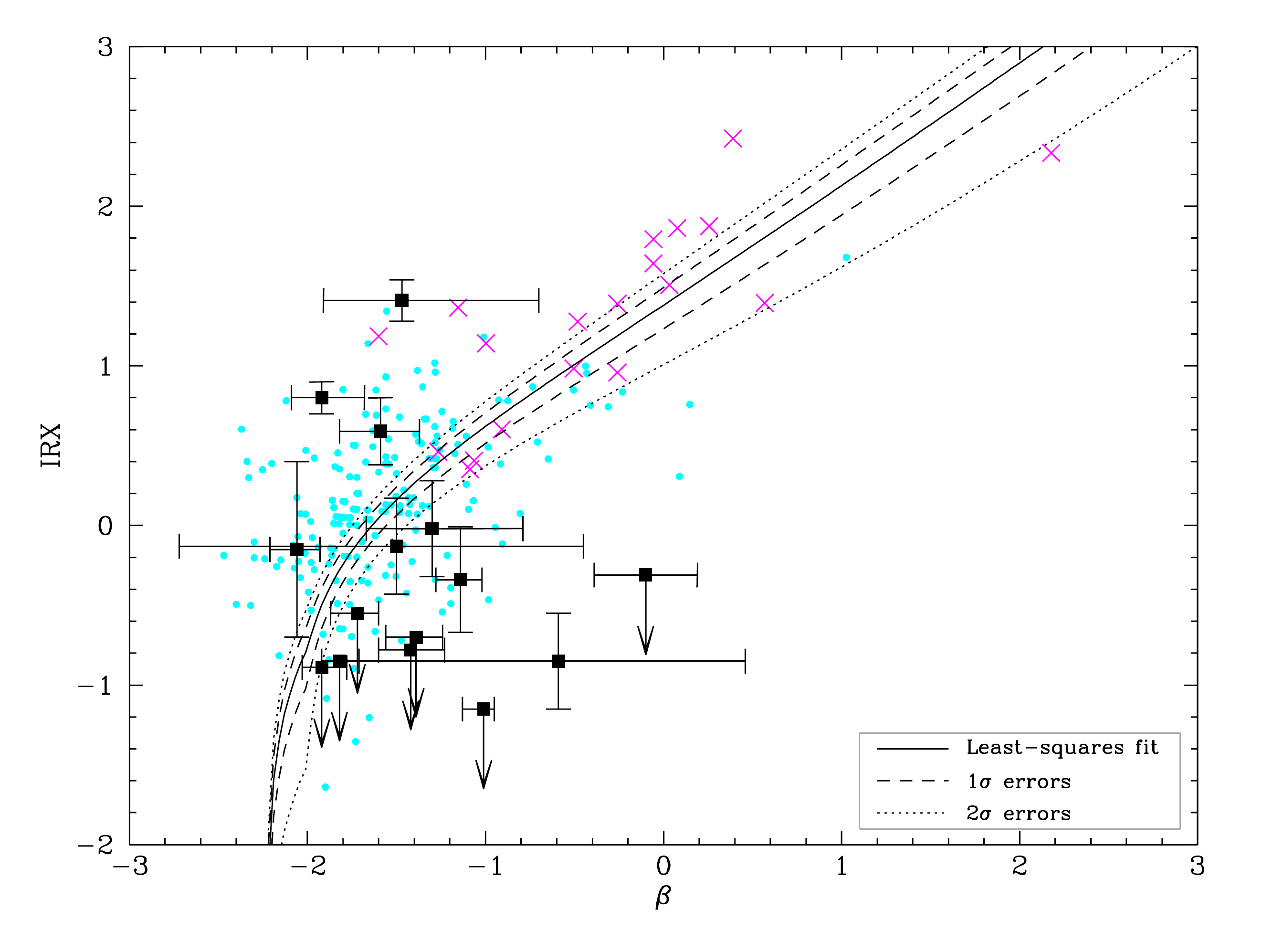}
\end{minipage}
\caption{{\itshape Left panel:} theoretical IRX-to-$\beta$ relations for mixed AGN and host-galaxy UV contributions. A $\beta_{\rm int}^{\rm (q)} = -2$ AGN dominance of 30\% ({\itshape solid line}) to 70\% ({\itshape dashed line}) has been adopted. For comparison, the theoretical IRX-to-$\beta$ relation by \citet{Meurer1999} with $\beta^{\rm (g)}_{\rm int} = -2.23$ used to evaluate the UV host-galaxy contribution ({\itshape dot-dashed line}) and the relation by \citet{Pettini1998} for the UV AGN contribution with $\beta^{\rm (q)}_{\rm int} = -2$ ({\itshape dotted line}) are also shown superimposed to the data sets by \citet[][{\itshape black squares}\ignorespaces ]{Barisic2017}, \citet[][{\itshape cyan dots}\ignorespaces ]{Dong2016} and \citet[][{\itshape magenta crosses}\ignorespaces ]{Garcia2016}. {\itshape Right panel:} least-squares fit of Eq. \ref{eqn:irxfq} ({\itshape solid line}) to the high-$z$ galaxy data by \citet{Barisic2017} and the AGN samples by \citet{Dong2016} and \citet{Garcia2016}, along with its boundaries at $1\sigma$ ({\itshape dashed lines}) and $2\sigma$ level ({\itshape dotted lines}).}
\label{fig:teorico}
\end{figure*}

\subsection{Estimate of the galactic UV contribution}

The scenario in which the high-$z$ galaxies by \citet{Capak2015} host AGN activity with average spectral properties (i.e., $\beta_{\rm int} \sim -2$ once a correction for the dust reddening is taken into account in evaluating the spectral slope; \citealt{Dong2016}) holds well in describing the observed IRX-to-$\beta$ distribution of such objects. However, a residual galaxy contribution to the total UV flux cannot be neglected, since models of AGN and host-galaxy coupling, derived from numerical simulations \citep[e.g.,][]{Schneider2015},  show that the AGN emission accounts on average for 30\% to 70\% of the system's UV energetics. Therefore, we tried to reproduce the observed values of IRX and $\beta$ for the high-$z$ galaxies with a hybrid relation that takes into account both the AGN and host-galaxy UV contribution.

Equation \ref{eqn:irx} can be rewritten in a form that only depends on the fraction $D(\tau_\lambda,\beta)$ of reprocessed UV light $L_{1600}$ into IR radiation and on the UV dust reddening $S(\tau_\lambda,\beta)$. In doing this, we obtain $L_{\rm UV} = L_{1600} \cdot S(\tau_\lambda,\beta)$, $L_{\rm FIR} = L_{1600} \cdot D(\tau_\lambda,\beta)$ and
\begin{equation}\label{eqn:irxds}
{\rm IRX} = \log{\left[
\frac{L_{\rm 1600}\cdot D(\tau_\lambda,\beta)}{L_{\rm 1600} \cdot S(\tau_\lambda,\beta)}
\right]}=\log{\left[
\frac{D(\tau_\lambda,\beta)}{S(\tau_\lambda,\beta)}
\right]}
\end{equation}
The reprocessed UV fraction $D(\tau_\lambda,\beta)$ and the reddening $S(\tau_\lambda,\beta)$ are respectively given by
\begin{equation}\label{eqn:dkb}
D(\tau_\lambda,\beta) = \int_{1500}^{3000} \left(\frac{\lambda}{1600 \mbox{ \AA}}\right)^\beta\left(1-e^{-\tau_\lambda}\right) d\lambda
\end{equation}
and
\begin{equation}\label{eqn:skb}
S(\tau_\lambda,\beta) = \int_{1500}^{3000} \left(\frac{\lambda}{1600 \mbox{ \AA}}\right)^{\beta}e^{-\tau_\lambda} d\lambda,
\end{equation}
where $\beta$ is the intrinsic slope of the UV spectrum and $\tau_\lambda$ is the wavelength-dependent optical depth of the dust distribution.

Since the AGN (through accretion-disk emission) and the host galaxy (through star formation) both heat the dust that reprocesses the UV into FIR emission, we can separate the contributions to the IRX by writing
\begin{equation}\label{eqn:irxqg}
{\rm IRX} = \log{\left\{
\frac{
                        L^{\rm (q)}_{\rm 1600}\cdot D\left[\tau^{\rm (q)}_\lambda,\beta_{\rm q}\right] + L^{\rm (g)}_{\rm 1600}\cdot D\left[\tau^{\rm (g)}_\lambda,\beta_{\rm g}\right] 
                        }{
                        L^{\rm (q)}_{\rm 1600}\cdot S\left[\tau^{\rm (q)}_\lambda,\beta_{\rm q}\right] + L^{\rm (g)}_{\rm 1600}\cdot S\left[\tau^{\rm (g)}_\lambda,\beta_{\rm g}\right]             
            }
            \right\}},
\end{equation}
where $\beta_{\rm q}$ and $\beta_{\rm g}$ are the AGN and host-galaxy UV spectral slopes respectively. Defining $f_{\rm q}$ as the fraction of energy injected in the dust by the AGN, we get
\begin{equation}\label{eqn:fracq}
f_{\rm q}=\frac{
                        L^{\rm (q)}_{\rm 1600}\cdot S\left[\tau^{\rm (q)}_\lambda,\beta_{\rm q}\right] 
                        }{
                        L^{\rm (q)}_{\rm 1600}\cdot S\left[\tau^{\rm (q)}_\lambda,\beta_{\rm q}\right] + L^{\rm (g)}_{\rm 1600}\cdot S\left[\tau^{\rm (g)}_\lambda,\beta_{\rm g}\right]             
            }
\end{equation}
and finally, combining Eqs. \ref{eqn:irxds}, \ref{eqn:irxqg}, and \ref{eqn:fracq},
\begin{equation}\label{eqn:irxfq}
        {\rm IRX} = \log{\left[
        f_{\rm q} \cdot 10^{{\rm IRX}_{\rm q}}+ \left(
        1-f_{\rm q}
        \right) \cdot 10^{{\rm IRX}_{\rm g}}
        \right]} 
\end{equation}

We adopt the functional form by \citet{Pettini1998} with $\beta_{\rm int} = -2$ in order to describe the ${\rm IRX}_{\rm q}$. This model is in fact valid if the dust distribution is foreground to the UV source \citep[e.g.,][]{Gordon1997,Inoue2005,Gallerani2010}, which is suggested for AGN by unified scenarios where a clumpy distribution of obscuring dust surrounds the central engine \citep[e.g.,][]{Urry1995,Risaliti2011,Krumpe2014}. For ${\rm IRX}_{\rm g}$, we adopt the \citet{Meurer1999} IRX-to-$\beta$ relation with $\beta_{\rm int} = -2.23$, since in galaxies the dust is expected to be mixed with stars rather than being (almost) completely foreground \citep{Calzetti1997,Pettini1998,Meurer1999}. In order to cover the whole average range of AGN contributions to the total UV emission, we choose for $f_{\rm q}$ the extreme values of 0.3 and 0.7 \citep{Schneider2015}.

Figure \ref{fig:teorico} shows such combined IRX-to-$\beta$ relations, along with the pure relations by \citet{Pettini1998} and \citet{Meurer1999}. The relations with 30\% to 70\% UV contribution from AGN emission do not fully describe any of the data by \citet{Barisic2017}, while the \citet{Meurer1999} relation partially intercepts the observations at IRX $\gtrsim 0$ and the \citet{Pettini1998} relation partially succeeds in describing the bulk of observations at $-1 <$ IRX $<0$ and $-2 < \beta < -1$. Overall, an IRX-to-$\beta$ relation with both an AGN and galactic contribution to the UV emission appears to be the most suitable to simultaneously describe the spectral properties of Seyferts, quasars, and star-forming galaxies at high redshift. The posterior distribution of least-squares fits of Eq. \ref{eqn:irxfq} performed over bootstrap realizations of the total data set by \citet[][high-$z$ galaxies with anomalous IRX]{Barisic2017}, \citet[][HerS/SDSS quasars]{Dong2016}, and \citet[][local Seyferts]{Garcia2016} with $f_{\rm q}$ as the only free parameter gives  $f_{\rm q} = 0.73 \pm 0.11$,  compared with a value of  $f_{\rm q} = 0.70 \pm 0.12$ obtained by fitting only the AGN-dominated objects.

\subsection{Intrinsic reddening estimates of the high-$z$ galaxies}

Under the assumption that the \citet{Capak2015} galaxies are actual observations of AGN hosts that exactly follow an IRX-to-$\beta$ relation with AGN contribution, it is now possible to derive the unattenuated AGN spectral slopes $\beta_{\rm int}^{\rm (q)}$ using the measured IRX and the observed UV slopes $\beta_{\rm obs}$. In order to obtain these values, we evaluate Eq. \ref{eqn:irxfq} for each high-$z$ galaxy with $f_{\rm q}$ between 0.3 and 1, fixing the intrinsic slope for the galactic contribution $\beta_{\rm int}^{\rm (g)}$ at $-2.23$ \citep{Meurer1999} and adopting as our fiducial estimate the value of $\beta_{\rm int}^{\rm (q)}$ for which the composite IRX-to-$\beta$ relation intercepts each data point. In particular, we find that the lowest AGN contribution (i.e., $f_{\rm q} = 0.3$) is the most reasonable choice to reproduce the observed IRX-to-$\beta$ values for the four leftmost objects in the IRX-to-$\beta$ diagram, namely HZ4, HZ9, HZ10, and HZ10W. This choice of $f_{\rm q}$ predicts values of $\beta_{\rm int}^{\rm (q)}$ between $-5.7$ and $-2.8$ for such objects; higher values of the AGN contribution would predict even steeper intrinsic AGN UV slopes or would fail to reproduce the observed data.

The remaining objects are instead well reproduced if we set a complete AGN domination in the UV flux (i.e., $f_{\rm q} = 1$). Lower values of $f_{\rm q}$ lead  the composite IRX-to-$\beta$ relation to intercept the region between the IRX-to-$\beta$ fiducial values below $\beta_{\rm int}^{\rm (q)} \sim -2$. An AGN-dominated UV spectrum allows us instead to reproduce the data of HZ1, HZ2, HZ3, HZ6b, and HZ6c with a single relation at $\beta_{\rm int}^{\rm (q)} = -2$; we recall that this is the average value of the intrinsic UV slopes for AGN \citep{Dong2016}. Furthermore, such a relation also describes the remaining objects assuming $-2 \lesssim \beta_{\rm int}^{\rm (q)} \lesssim 0$.

We then use the $\beta_{\rm int}$ derived in this way along with the {\itshape HST}/WFC3 data by \citet{Barisic2017} in order to obtain an estimate of the UV extinction $A_{1600}$ affecting the AGN emission. To do this, we adopt the simplified assumption that the bolometric luminosities of these objects are  approximated using the optical/UV bolometric corrections by \citet{Runnoe2012}. Therefore, we estimate the optical monochromatic luminosities at 3000 \AA~from the F160W magnitudes by \citet{Barisic2017}, and apply the corrections listed in Table 1 of \citet{Runnoe2012c} to derive the de-reddened UV luminosities at 1600 \AA. The choice of indirectly computing such luminosities through the correction of the optical magnitudes is motivated by the fact that, at least for low to moderate values of $A_{1600}$, the luminosity at optical wavelengths is less affected by extinction than the UV flux.

\begin{figure}
\includegraphics[width=.75\columnwidth,angle=-90]{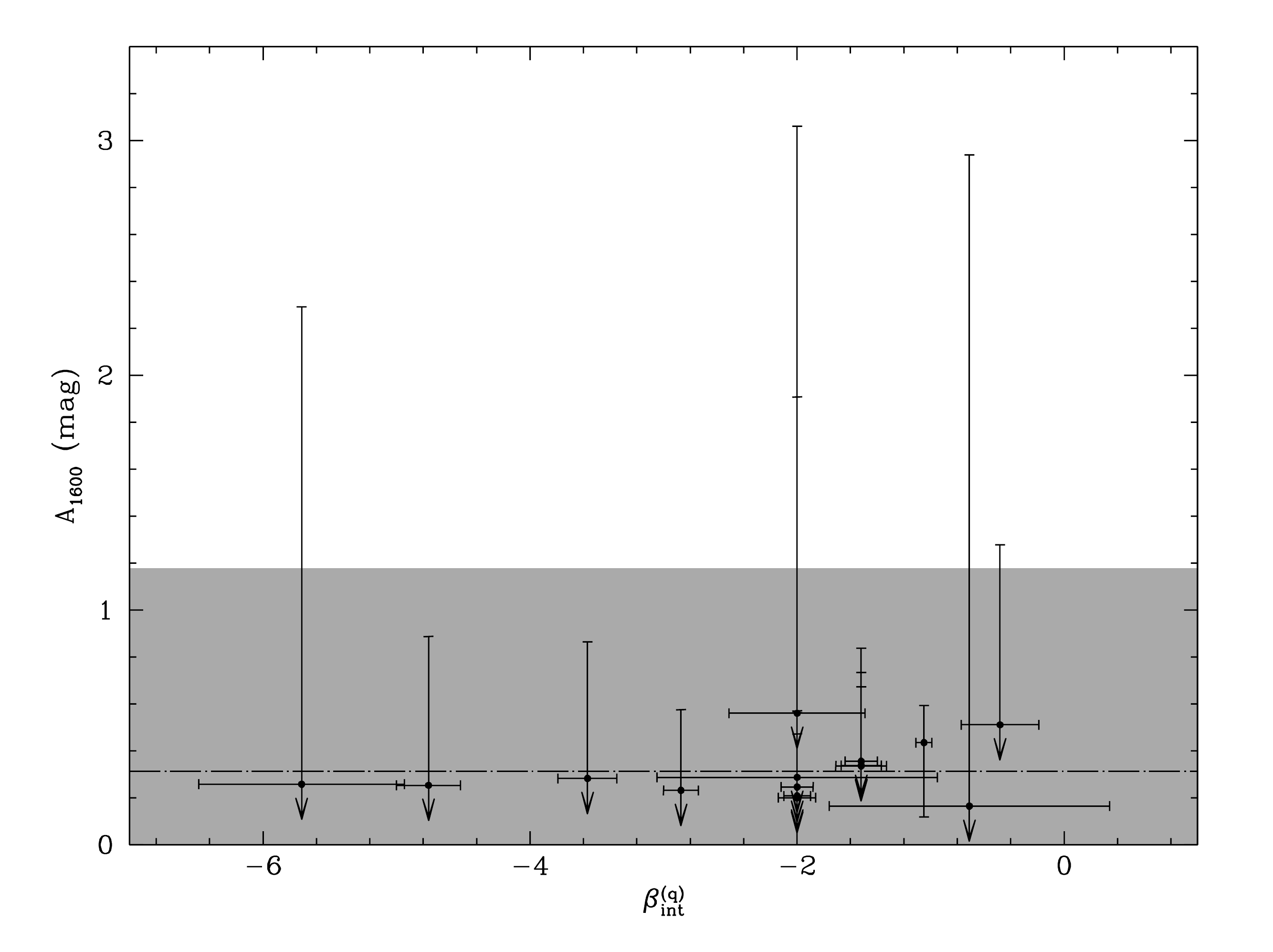}
\caption{Extinction coefficient $A_{1600}$ as a function of the intrinsic spectral slope $\beta_{\rm int}^{\rm (q)}$ for the high-$z$ galaxies by \citet{Capak2015}. The average extinction ({\itshape dot-dashed line}) is indicated along with its 1$\sigma$ confidence interval ({\itshape gray band}).}
\label{fig:cabi}
\end{figure}

Under these hypotheses, we define the values of $A_{1600}$ by assuming that the intrinsic UV slopes of the objects by \citet{Capak2015} are reddened by an SMC-like extinction. Such extinction coefficients are shown in Fig. \ref{fig:cabi} as a function of $\beta_{\rm int}^{\rm (q)}$, suggesting that the possible AGN emission of these high-$z$ galaxies is affected on average by low to moderate reddening along the line of sight ($A_{1600} \lesssim 1.2$ mag).

\subsection{Possible issues in the comparison between different samples of extragalactic sources}\label{sec:caveats}

Although the presented analysis focuses on the comparison among the average emission properties of the sample of high-$z$ star-forming galaxies by \citet{Capak2015} with those of two AGN classes, the dependence of the latter on quantities derived from SED fitting can bias the results due to (i) the availability of different models for the paramount spectral components , (ii) the possible parameter degeneracies arising in the SED construction, and (iii) the adoption of peculiar extinction curves. In the following, we briefly describe the expected impact of these biases on the main findings reported in this section.

\paragraph*{{\itshape Model dependencies.}} {Model dependencies are particularly important in fitting the main sources of UV (e.g., the details of the AGN process and its ionized surrounding environment) and IR emission (e.g., the reprocessing of radiation by dust in the AGN's surrounding environment). For instance, the AGN emission may be reproduced by either using a UV template to take into account the additional light from resolved emission lines and Fe {\scriptsize II} blending \citep{Garcia2016} or adopting a simple power-law functional form \citep{Dong2016}. However, the difference in such approaches generates discrepancies of $\lesssim$10\% only in the determination of the UV flux from AGN, i.e., the average contribution of emission features to the total emission with respect to the continuum \citep[e.g.,][]{Vandenberk2001,Glikman2006}, and hence is not expected to have a great impact on the derivation of the source's properties unless tailoring the selection for objects with extreme line emission. On broader wavelength ranges, the differences introduced by SEDs related to different emission models can directly affect the determination of IRX, UV slope, and dust extinction. \citet{Duras2017} show that there are some differences among quasar SEDs, but these are in general compatible with each other within at most 0.5 dex on a wavelength range of 100 to $10^6$ \AA\ (see, e.g., their figures 2 and 3), apart from some peculiar choices \citep[e.g.,][]{Stalevski2016}. Additionally, the largest discrepancies in the IR emission are due to the opening angle of a dusty AGN torus \citep{Stalevski2016,Duras2017}, but they are again limited to $\sim$0.5 dex. A shift of $\sim \pm 1$ dex in the IRX of AGN introduced by SED differences is still compatible with the observed IRXs of the zombie galaxies by \citet{Capak2015}, thus not greatly biasing our main results.}

\paragraph*{{\itshape Degeneracies.}} {Another problem is the degeneracies among the free parameters involved in the SED-fitting procedure, which can be explored in detail only with the availability of rich data sets in order to give statistical significance to the fit results  \citep[see, e.g.,][]{Mullaney2011}. For instance, the UV SED of a quasar affected by intrinsic reddening could be reproduced by either assuming a shallow UV slope (thus neglecting the dust extinction) or an appropriate extinction law applied to a steep spectrum \citep[$\beta \sim -2$; e.g.,][]{Dong2016}. In the FIR, both the dust emission from the host galaxy and the emission from a possible old stellar population peak at $\sim$100 $\mu$m, thus making it difficult to disentangle  quasar and galaxy light. \citet{Dong2016} overcome these issues by adopting quasar-galaxy mixing diagrams \citep{Hao2013} that allow us to distinguish AGN-dominated sources from galaxy-dominated and reddened SEDs. Additionally, they also fix some SED parameters to typical literature values, such as the AGN torus temperature at 1300 K and the slope of the graybody emissivity for galactic dust at 1.6 in order to reduce possible secondary degeneracies. On the other hand, \citet{Garcia2016} rely on spectro-photometric indicators such as the $f_\nu(\mbox{70 }\mu{\rm m})/f_\nu(\mbox{160 }\mu{\rm m})$ flux ratio and the equivalent width of polycyclic aromatic hydrocarbon (PAH) features to discriminate between galaxy emission and AGN activity. Therefore, we can assume that parameter degeneracies are already taken into account in the adopted AGN samples, thus not affecting their comparison with the properties of the zombie galaxies.}

\paragraph*{{\itshape Dust extinction.}} {Regarding the choice of the extinction curve, we have already outlined that taking into account different extinction curves and dust distribution geometries do not help  reconcile the expected behavior of the zombie galaxies in the IRX--$\beta$ plane, as widely described in the literature \citep[see, e.g.,][and references therein]{Mancini2016, Narayanan2017, Ferrara2017, Behrens2018}. In particular, the work by \citet{Narayanan2017} suggests that in order to have a galaxy that covers the same locus of the zombie galaxies these object would need a spectral emission dominated by an old stellar population. Alternatively, \citet{Ferrara2017} suggest that it is possible to reduce the discrepancy between the observed and expected IRXs by considering a stellar population completely embedded in a molecular cloud phase. However, even in this strong condition, it is not possible to completely explain the FIR upper limits measured by \citet{Capak2015} and \citet{Barisic2017}. Another possibility is to consider a highly inhomogeneous dust distribution produced by dust accretion of surface metals \citep[e.g.,][]{Mancini2016}, which would produce higher values of $\beta$ at the same IRXs. Nevertheless, this effect is not strong enough to reconcile the theoretical expectations of UV slopes and FIR emission with the observed behavior of the zombie galaxies, particularly the upper limits in the IRX at $\beta \gtrsim -1.5$ \citep[see][for further details]{Mancini2016}. Therefore, we remark that the existence of a weak AGN activity at the center of such high-$z$ objects appears to be a more reasonable explanation for their peculiar IRX-to-$\beta$ values.}

\section{Color-magnitude relation for AGN}\label{sec:cmragn}

The difficulty to reproduce the IRX-to-$\beta$ relation for the sources by \citet{Capak2015} within the scenario of galaxy-dominated objects can be avoided by assuming that these sources are actually hosting AGN activity with UV spectral emission depending on the details of the accretion process. Figures \ref{fig:irxcolored}, \ref{fig:IRXSy}, and \ref{fig:teorico} show  how a UV spectrum with $\beta_{\rm int}$ not fixed to galactic values may explain the bulk of the data after the reanalysis by \citet{Barisic2017}. If this alternative scenario holds, these objects should also follow typical relations that are known for AGN, such as showing steeper spectral slopes at higher luminosities \citep[e.g.,][]{Shankar2016}.

\begin{figure}
\includegraphics[width=\columnwidth]{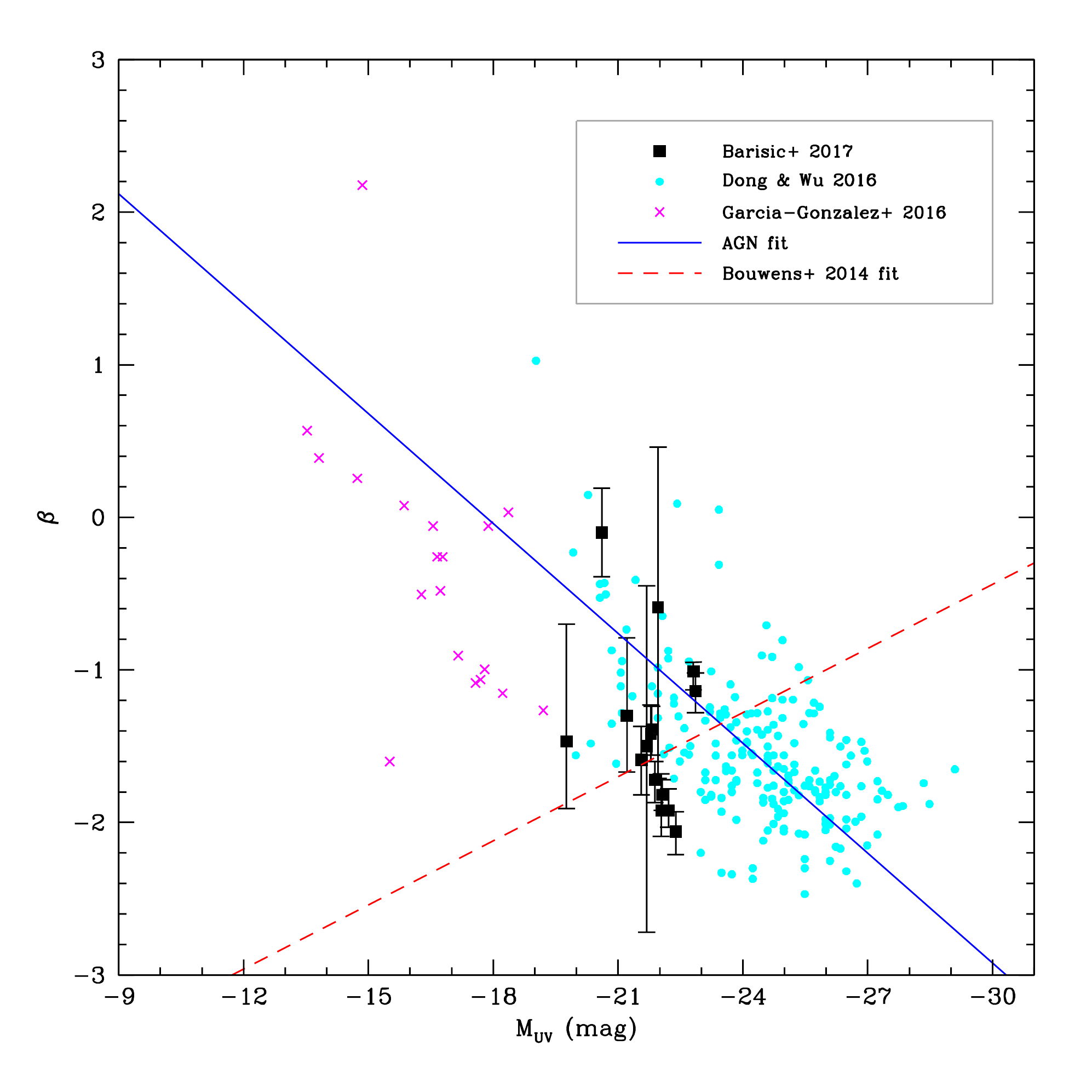}
\caption{Color-magnitude relation for the HerS/SDSS quasars by \citet[][{\itshape cyan dots}\ignorespaces ]{Dong2016}, the Seyfert galaxies by \citet[][{\itshape magenta crosses}\ignorespaces ]{Garcia2016}, and the high-$z$ galaxies by \citet[][{\itshape black squares}\ignorespaces ]{Barisic2017}. The $\beta$-to-$M_{\rm UV}$ relation derived by fitting the HerS/SDSS quasars ({\itshape blue solid line}) is shown superimposed to the data. For comparison, the relation valid for galaxies at $z \sim 5$ \citep[][{\itshape red dashed line}\ignorespaces ]{Bouwens2014} is also shown.}
\label{fig:cmr}
\end{figure}

Figure \ref{fig:cmr} shows $\beta$ as a function of $M_{\rm UV}$ for the HerS/SDSS quasars, the Seyfert galaxies by \citet{Garcia2016}, and the high-$z$ galaxies by \citet{Capak2015}. In order to quantify the $\beta$-to-$M_{\rm UV}$ relation for AGN, we perform a Monte Carlo linear regression by running the script {\scriptsize PyMC3} \citep{Salvatier2016} on the joint sample of quasar and Seyfert spectral slopes $\beta$ as a function of their UV absolute magnitude $M_{\rm UV}$. We choose a standard likelihood defined as
\begin{equation}
\mathcal{L} \propto \exp{\left(
-\frac{\chi^2}{2}
\right)} = \exp{\left\{
- \frac{\left[
\beta -(a\cdot M_{\rm UV}+b)
\right]^2}{2 \sigma^2}
\right\} }
\end{equation} 
with very broad priors for the parameters $a$, $b$, and $\sigma$:
\begin{equation}
\begin{array}{lr}
a \sim  \mathcal{N}_{(0,10)}, \\
b \sim \mathcal{N}_{(0,10)},\\
\sigma \sim \mathcal{N^{+}}_{(0,1)},\\
\end{array}
\end{equation}
where $\mathcal{N}_{(\mu,k)}$ is the Gaussian distribution of UV magnitudes with a given average $\mu$ and variance $k^2$, and $N^+_{(\mu,k)}$ is its positive side. In this way, we find
\begin{equation}
\beta_{\rm AGN}  = \left(
0.24 \pm 0.02
\right)\cdot \left( M_{\rm UV} +19.5 \right) - \left(
0.47 \pm 0.03 \pm 0.14
\right),
\label{eq:cmrdong}
\end{equation}
where we adopt the average uncertainty in the measurement of $\beta$ for HerS/SDSS objects as an estimate of the systematic error on the intercept of the color-magnitude relation.

\begin{table}
\centering
\resizebox{\columnwidth}{!}{
\begin{tabular}{lccc}
\hline 
\hline
ID & $\chi_{\rm AGN}^2$ & $\chi_{\rm gal}^2$ & $P_{\rm AGN}$\\
\hline
HZ1 &  9.663  &  4.481 & 0.07\\
HZ2 &  9.175  &  2.649 & 0.04\\
HZ3 &  6.967  &  0.597 & 0.04\\
HZ4 &  11.25  &  7.939 & 0.16\\
HZ5 &  1.001  &  6.270 & 0.93\\
HZ5a &  --  &  -- & --\\
HZ6 &  0.255  &  2.319 & 0.74\\
HZ6a &  0.181  &  0.796 & 0.58\\
HZ6b &  0.190  &  0.008 & 0.48\\
HZ6c &  0.732  &  0.649 & 0.49\\
HZ7 &  1.842  &  0.927 & 0.39\\
HZ8 &  1.994  &  0.577 & 0.33\\
HZ8W &  3.348  &  28.95 & 1.00\\
HZ9 &  4.105  &  0.015 & 0.11\\
HZ10 &  7.697  &  2.208 & 0.06\\
HZ10W &  2.073  &  0.406 & 0.30\\
\hline
\end{tabular}
}
\caption{Values of  $\chi^2$ obtained by comparing the \citet{Barisic2017} data with Eq. \ref{eq:cmrdong} and \ref{eq:bouwens-fit}, respectively, along with the corresponding Bayesian probabilities of AGN domination (see text for details).}
\label{tab:probs}
\end{table}

Conversely, the dependence of $\beta$ on UV emission in star-forming galaxies is mainly determined by the amount of extinction. Therefore, the expected trend is to find shallower spectral slopes in bigger, more luminous, more chemically enriched objects with respect to fainter ones \citep[e.g.,][]{Meurer1999,Bouwens2014,Finkelstein2015,Finkelstein2015b}. Here we adopt the relation inferred by \citet{Bouwens2014} at $z \sim 5$ to describe the relation between $\beta$ and $M_{\rm UV}$ in galaxy-dominated objects,
\begin{equation}
\beta_{\rm gal} =  -(0.14 \pm 0.02)\cdot\left(M_{\rm UV}+19.5\right) -(1.91 \pm 0.02 \pm 0.06),
\label{eq:bouwens-fit}
\end{equation}
which exhibits an opposite trend with respect to the relation for AGN.

The sample of objects by \citet{Capak2015} is numerically small if compared to the HerS/SDSS ensemble (207 objects) or the \citet{Bouwens2014} catalog ($>$4000 objects). Furthermore, a visual inspection of Fig. \ref{fig:cmr} reveals that this sample tends to cluster around the $\beta$-to-$M_{\rm UV}$ relation for high-$z$ galaxies. Nevertheless, we aim to test statistically the probability that these objects are AGN- or galaxy-dominated in terms of relative proximity to one relation with respect to the other. In order to do this, we define a $\chi^2$ variable as
\begin{equation}\label{eq:chi2}
\chi^2_{\rm x} = \frac{\left(
\beta - \beta_{\rm x}
\right)^2}{\sigma_{\beta}^2 + \sigma_{\rm x}^2},
\end{equation}
where $\beta$ and $\sigma_{\beta}$ refer to the observed slopes with their average errors by \citet{Barisic2017}, while $\beta_{\rm x}$ are the values of UV spectral slopes expected for the \citet{Capak2015} objects using either Eq. \ref{eq:cmrdong} ($\beta_{\rm x} = \beta_{\rm AGN}$) or Eq. \ref{eq:bouwens-fit} ($\beta_{\rm x} = \beta_{\rm gal}$) with associated standard deviations $\sigma_{\rm x}$. The Bayesian probability $P_{\rm AGN}$ for a data point by \citet{Barisic2017} to come from an AGN-dominated object is then
\begin{equation}\label{eq:pagn}
P_{\rm AGN} = \frac{1}{1 + \exp{\left(
\Delta\chi^2/2
\right)}},
\end{equation}
with $\Delta\chi^2 = \chi^2_{\rm AGN} - \chi^2_{\rm gal}$ for an object having $\chi^2_{\rm AGN}$ with respect to Eq. \ref{eq:cmrdong} and $\chi^2_{\rm gal}$ with respect to Eq. \ref{eq:bouwens-fit}. The complementary probability $P_{\rm gal} = 1 - P_{\rm AGN}$ is thus the probability  that the data point  represents a galaxy-dominated object. We note that we  assume a flat prior on both relations (i.e., the objects by \citealt{Capak2015} are not forced to follow {a priori} one between Eq. \ref{eq:cmrdong} or \ref{eq:bouwens-fit}).

We report the probabilities in Table \ref{tab:probs} for each of the high-$z$ objects by \citet{Barisic2017}. Only HZ5 and HZ8W can be classified as AGN hosts at a 90\% confidence level, with HZ5 already recognized as an X-ray selected quasar \citep{Capak2015}, whereas HZ1 to HZ3 and HZ10 are likely to be normal galaxies. The other targets show intermediate values of $P_{\rm AGN}$, and hence it is not possible to strictly classify them in one category, though all of the objects are enclosed within the scatter of the joint AGN color-magnitude relation. Therefore, no firm conclusion may be drawn about the relative spectral domination of AGN or galaxy emission in the sample of objects by \citet{Capak2015}. In order to better distinguish AGN-dominated objects from galaxies, we define as $\Delta\beta$ the distance of the observed $\beta$ from the $\beta$-to-$M_{\rm UV}$ relation for AGN described by Eq. \ref{eq:cmrdong},
\begin{equation}
\Delta\beta = \beta^{i} -0.24\cdot \left( M^{i}_{\rm UV} +19.5 \right) + 0.47,
\end{equation}
where $\beta^{i}$ and $M^{i}_{\rm UV}$ are the UV spectral slope and absolute magnitude for the $i$-th object.

\begin{figure}
\includegraphics[width=\columnwidth]{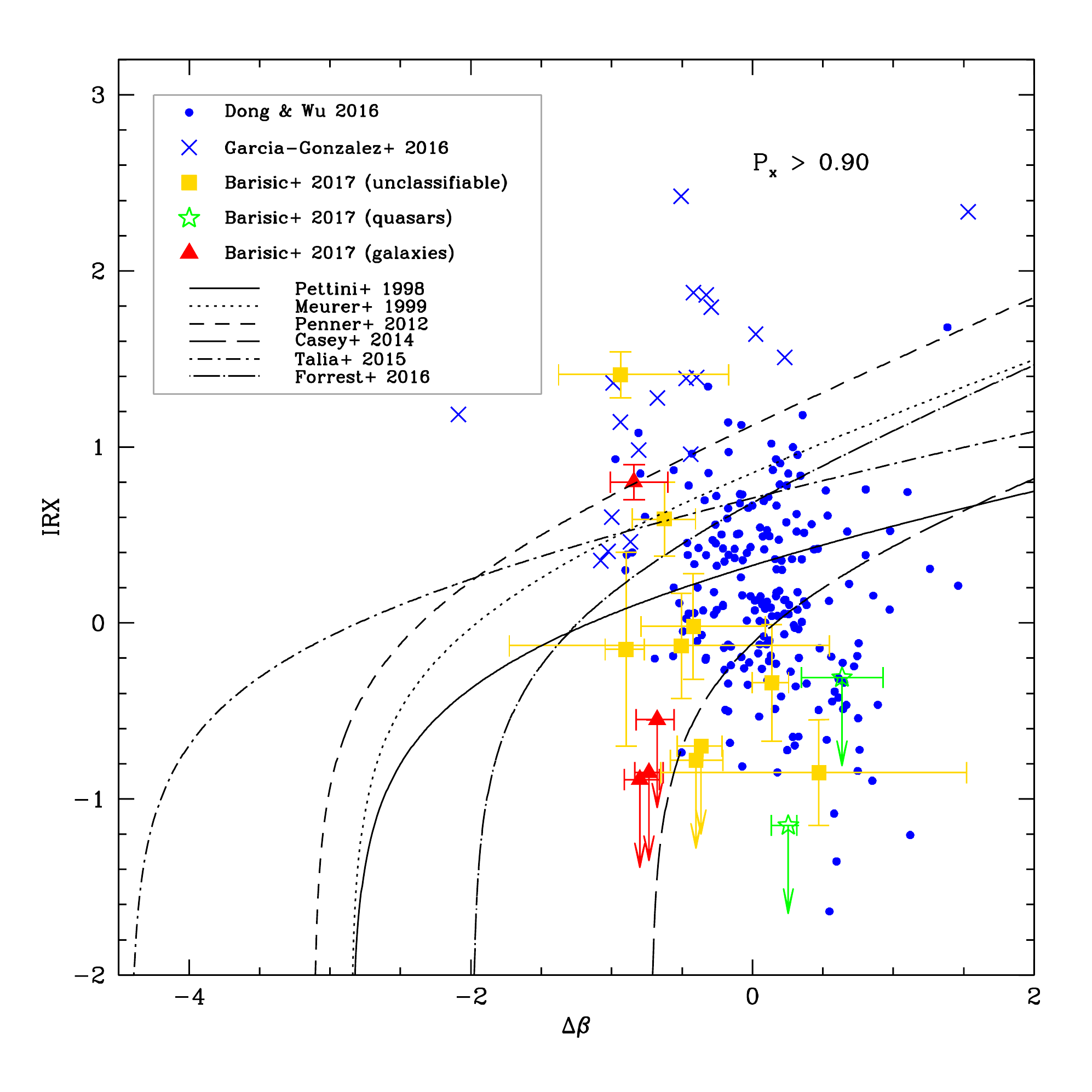}
\caption{ IRX as a function of the slope difference $\Delta\beta$ for the HerS/SDSS quasars by \citet[][{\itshape blue dots}\ignorespaces ]{Dong2016}, the Seyfert galaxies by \citet[][{\itshape blue crosses}\ignorespaces ]{Garcia2016}, and the high-$z$ galaxies by \citet{Barisic2017}. The high-$z$ galaxies are classified according to the statistical analysis presented in Sect. \ref{sec:cmragn} ({\itshape red triangles:} actual galaxies, {\itshape yellow squares:} unclassifiable objects, {\itshape green stars:} actual quasars). The curves available in the literature that describe the IRX-to-$\Delta\beta$ relation for various classes of galaxies \citep{Pettini1998,Meurer1999,Penner2012,Casey2014,Talia2015,Forrest2016} are also shown.}
\label{fig:irxfunc}
\end{figure}

We  show this quantity in Fig. \ref{fig:irxfunc} as a function of the IRX for the \citet{Dong2016}, the \citet{Garcia2016} and the \citet{Barisic2017} data. It can be seen at  first glance that the AGN are clustered by construction around a value of $\Delta\beta \sim 0$, and scattered over a range of IRX $\in [-2,3]$ with the bulk of the data comprised between IRX $\sim -1$ and $2$. The objects from \citet{Capak2015} are superimposed on the region covered by actual AGN, with the tendency to populate the lower part of the diagram. Interestingly, segregating these objects by their classification based on Bayesian probability shows that the galaxy-dominated objects (i.e., those objects with $P_{\rm gal} > 0.90$) cluster around $\Delta\beta \sim -0.8$, while the AGN hosts exhibit positive values of $\Delta\beta$ up to $\sim$1. This diagram synthesizes all the available information about the emission properties of the \citet{Capak2015} galaxies as reanalyzed by \citet{Barisic2017} since it correlates $M_{\rm UV}$ with the $\rm IRX$ and $\beta_{\rm obs}$. We note that this kind of analysis could be used for a preliminary discrimination between AGN hosts and normal star-forming galaxies where a measurement of the FIR emission is feasible, although a more thorough analysis of their UV and FIR properties is needed. Furthermore, the IRX--$\Delta\beta$ plane might also be useful for analyzing the evolution of the local galaxies.

\begin{table}
\label{Table:coeffs}
\centering
\resizebox{\columnwidth}{!}{
\begin{tabular}{lccr}
\hline
\hline
Paper & $m$ & $q$ & Sample\\
\hline
\citet{Pettini1998} & $0.39$ & $1.11$ & $z \sim 3$ Ly-break galaxies\\
\citet{Meurer1999} & $0.74$ & $2.11$ & Local starbursts\\
\citet{Penner2012} & $0.87$ & $2.71$ & $z \sim 2$ DOGs\\
\citet{Casey2014} & $0.75$ & $0.54$ & Local low-IRX galaxies\\
\citet{Talia2015} & $0.41$ & $1.81$ & $1 < z < 3$ SFG spectra\\
\citet{Forrest2016} & $0.88$ & $1.75$ & $1 < z < 3$ composite SEDs\\
\hline
\end{tabular}
}
\caption{Coefficients of the IRX-to-$\Delta\beta$ relations specialized from  literature results on different samples of galaxies.}
\end{table}

To directly compare the effect of removing the AGN dependence on $M_{\rm UV}$ from samples of different objects (quasars, Seyferts, local starbursts, medium-$z$ star-forming galaxies), we rewrite the IRX-to-$\beta$ relations available in the literature as a function of $\Delta\beta = \beta_{\rm gal} - \beta_{\rm AGN}$. Substituting Eq. \ref{eq:bouwens-fit} into Eq. \ref{eq:cmrdong} to eliminate the dependence on $M_{\rm UV}$ and then inverting the result to write $\beta_{\rm gal}$ as a function of $\Delta\beta$, we find
\begin{equation}\label{eqn:bgaldb}
\Delta\beta = 2.71 \beta_{\rm gal} + 3.74
\end{equation}
Therefore, the IRX-to-$\Delta\beta$ relations assume the form
\begin{equation}\label{eqn:irxdb}
{\rm IRX}\left(
\Delta\beta
\right) = \log{\left[
10^{0.4 \left(
m \cdot \Delta\beta + q
\right)}-1
\right]}+0.076,
\end{equation}
with the coefficients $m$ and $q$ depending on the analyzed sample of objects. Unlike in the previous sections of this paper, here we do not restrict ourselves to the relations by \citet{Pettini1998} and \citet{Meurer1999}. Instead, we focus on several relations available in the literature that are valid for samples of different classes of galaxies. In Table \ref{Table:coeffs}, we list the values of the coefficients $m$ and $q$ for the IRX-to-$\Delta\beta$ relations, along with the source paper of the original IRX-to-$\beta$ relation and a synthetic description of the corresponding type of astrophysical objects.

Figure \ref{fig:irxfunc} shows these IRX-to-$\Delta\beta$ relations superimposed on the AGN samples and on the high-$z$ galaxies by \citet{Capak2015}. Apart from the curve corresponding to the IRX-to-$\beta$ relation derived by \citet{Casey2014} for star-forming galaxies at $z<0.085$, it
is evident that none of the considered relations is able to intercept the \citet{Capak2015} galaxies with $\rm IRX<0$. This further strengthens the AGN-host hypothesis for the high-$z$ galaxies with anomalous IRXs, and again suggests that deepening the investigation of the IRX--$\Delta\beta$ plane should be  a future tool for object discrimination.

\begin{table*}
\begin{center}
\resizebox{0.8\textwidth}{!}{
\begin{tabular}{lcccc}
\hline 
\hline
ID & $\log{\left(L_{\rm UV}/L_\odot\right)}$ & SFR ($M_\odot$/yr) & $\log{\left(L_{\rm X}/L_\odot\right)}$ & $\log{\left(L^\star_{\rm X}/L_\odot\right)}$ \\
\hline 
HZ1 & 11.21 & 24 & 10.31 & 8.19 \\
HZ2 & 11.15 & 25 & 10.26 & 8.20 \\
HZ3 & 11.08 & 18 & 10.21 & 8.10 \\
HZ4 & 11.28 & 51 &10.36 & 8.38 \\
HZ5$^*$ & 11.45 & $<$3 & 10.82$^{**}$ & $<$7.59 \\
HZ5a & $<$10.37 & $<$3 & $<$9.66 & $<$7.59 \\
HZ6 & 11.47 & 49 & 10.50 & 8.35 \\
HZ6a & 11.11 & --- & 10.23 & ---  \\
HZ6b & 11.0 & --- & 10.15 & --- \\
HZ6c & 11.81 & --- & 10.00 & ---  \\
HZ7 & 11.05 & 21& 10.19 & 8.11 \\
HZ8 & 11.04 & 18& 10.18 & 8.06 \\
HZ8W & 10.57 & 6& 9.82 & 7.76 \\
HZ9 & 10.95 & 67& 10.11 & 8.46 \\
HZ10$^+$ & 11.14 & 169& 10.25 & 8.72 \\
HZ10W$^+$ & 10.23 & --- & 9.54 & ---  \\
\hline
\multicolumn{5}{l}{$^*$No [C {\scriptsize II}] redshift measurement available \citep{Capak2015}, photometric redshift is used.}\\
\multicolumn{5}{l}{$^{**}$Luminosity in the [0.5--2] keV energy range observed in the {\itshape C}-COSMOS field.}\\
\multicolumn{5}{l}{$^+$These objects fall outside the {\itshape C}-COSMOS field.}\\
\end{tabular} }
\caption{Luminosity properties of the objects described by \citet{Capak2015}. The expected X-ray luminosities in the observer-frame soft band have been estimated by adopting a bolometric correction to the UV luminosity of 5.2 \citep{Runnoe2012}.
\label{Table:Xray}
}
\end{center}
\end{table*}

\section{Expected X-ray luminosities}\label{sec:xraylum}

We now present the analysis of the X-ray emission expected for these objects within the AGN-host scenario. In fact, it has been clearly established that the accretion process around massive black holes can power the emission of X-rays from structures (jets, coronae) collateral to the ionized disk \citep[e.g.,][]{Halpern1982,Begelman1983a,Begelman1983b,Zdziarski1985}. Therefore, it is possible to select and characterize AGN by studying the details of their high-energy spectra \citep[e.g.,][]{Giustini2008,Lusso2010,Lusso2016,Page2017}.  In order to give a complete view of the X-ray energetics of the \citet{Capak2015} galaxies, we both evaluate the expected X-ray fluxes of such objects from their measured UV-to-IR properties and perform a stacking analysis on the {\itshape Chandra} X-ray imaging of the COSMOS field at the relevant sky coordinates.

\subsection{Estimated upper limits on the X-ray fluxes}

The field of the \citet{Capak2015} galaxies has been observed by the {\itshape Chandra} X-ray observatory in the framework of the {\itshape Chandra} COSMOS Survey\footnote{An extension of the {\itshape C}-COSMOS program, the 4.6 Ms {\itshape Chandra} COSMOS Legacy Survey \citep[e.g.,][]{Marchesi2016}, is present in the literature, but its imaging data are not publicly available for stacking analysis.} \citep[{\itshape C}-COSMOS;][]{Elvis2009,Puccetti2009,Civano2012,Lanzuisi2013}. This survey has imaged for 1.8 Ms the central region of the COSMOS field around 10$^h$ +02$^\circ$ with an average flux limit of $7.3 \cdot 10^{-16}$ erg s$^{-1}$ cm$^{-2}$ in the full band ($0.5 - 10$ keV). Of the objects listed in Table \ref{table:data}, HZ10 and HZ10W fall outside the area of 0.9 deg$^2$ covered by the {\itshape C}-COSMOS program, and the weak quasar HZ5 is detected as a resolved source. None of the other galaxies in the  \citet{Capak2015} sample have been detected. 

Assuming that the emission of the non-detected sources in the X-ray comes from AGN-dominated objects, we follow \citet{Runnoe2012} in computing the bolometric luminosities $L_{\rm bol}$ of such sources from the values of $L_{\rm UV}$ reported in Table \ref{table:data}. Therefore, we adopt  $L_{\rm bol} = f {\alpha_{\rm UV}} L_{\rm UV}$, where the UV-to-bolometric correction is $\alpha_{\rm UV} = 5.2$, and $f = 0.75$ is a factor taking into account the anisotropic emission from an accretion disk. The luminosities in the hard ($2-10$ keV) rest-frame X-ray band $L_{\rm X}$ can then be inferred from the relation obtained by \citet{Marconi2004},
\begin{equation}\label{eqn:mar04}
\log\left(
\frac{L_{\rm bol}}{L_{\rm X}}
\right)= 1.54 + 0.24\mathfrak{L} + 0.012\mathfrak{L}^2 - 0.0015\mathfrak{L}^3 , 
\end{equation}
where $\mathfrak{L} = \log(L_{\rm bol}) - 12$. These X-ray luminosity evaluations must be interpreted as upper limits on the intrinsic energetic output of these objects since they would overestimate the real values in the case of non-negligible contribution of the host-galaxy emission in the UV-to-optical band.

Nevertheless,  a significant fraction of the observed X-ray emission, especially at low X-ray luminosities ($L_{\rm X} \sim 10^{43}$ erg s$^{-1}$), may be powered by the cumulative radiation produced by high-mass X-ray binaries (HMXBs) in the host galaxy \citep{Ranalli2003, Ranalli2005, Lehmer2010, Lehmer2016}. Therefore, we estimate this contribution $L^\star_{\rm X}$ in the hard X-ray band following \citet{Volonteri2017}, and adopting the redshift-dependent relation found by \citet{Lehmer2016} for galaxies in the redshift range $z \in [0,7]$
\begin{equation}\label{eqn:hmxb}
\log\left(
L^\star_{\rm X}
\right) = 39.82 + 0.63\log({\rm SFR}) + 1.31 \log(1 + z),
\end{equation}
where the selected values for $z$ and SFR are shown in Table \ref{table:data} and Table \ref{Table:Xray},               respectively.

We report the predicted luminosities $L_{\rm X}$ and $L^\star_{\rm X}$ for the analyzed sample in Table \ref{Table:Xray}. It is clearly visible that, on average, the foreseen luminosities assuming an AGN-driven X-ray emission are $\sim$2 orders of magnitude higher than the values obtained under the hypothesis of X-ray flux produced by an HMXB population. Even accounting for a host-galaxy contribution of $\sim$70\%, the two predicted emissions still remain very different.

We further note that, at high redshifts, X-ray fluxes can be absorbed by highly obscured environments, such as large amounts of gas and dust surrounding accreting BHs \citep{Ueda2003, Gilli2007, Treister2009}. The two main attenuation processes in the X-rays are the photoelectric absorption, accounted for by the cross section $\sigma_{\rm ph}$, and the Compton scattering of photons against free electrons, $\sigma_{\rm T}$. The photoelectric cross section decreases with increasing energy, and the contribution of the dust to $\sigma_{\rm ph}$ is relatively modest for $E\gtrsim 1$ keV \citep{Morrison1983}. Therefore, at higher energies ($E \gtrsim 10$ keV for gas metallicities $Z \sim Z_\odot$), the energy-independent $\sigma_{\rm T}$ becomes dominant. Softer X-ray photons are thus expected to be more absorbed than harder ones. Furthermore, X-ray observations of AGN at $z \gtrsim 4$ typically sample the rest-frame hard X-ray emission, i.e., in the $3 - 12$ keV range on average for the sample by \citet{Capak2015}. Even assuming a strong obscuration produced by gas column densities $N_{\rm H} \sim 10^{23}$ cm$^2$, the unabsorbed AGN SED for $E \gtrsim 3$ keV and $Z \lesssim 0.1 Z_\odot$ is at most 1.2 times higher than the absorbed one \citep[see, e.g., fig. 3 in][]{Pezzulli2017}. The theoretical simulations \citep{Mancini2016} and observations \citep{Mannucci2010,Ma2016} both suggest that the expected metallicity for this kind of objects is  $\lesssim 0.1 Z_\odot$. This suggests that neither absorption nor a dominant contribution of the stellar emission to the observed optical photometry is able to reduce  the predicted AGN-driven X-ray emission enough to make $L_{\rm X}$ and $L^\star_{\rm X}$ comparable. Deeper X-ray observations may greatly help to discriminate between zombie and active galaxies.

\begin{figure*}
\begin{minipage}{.49\textwidth}
\centering
\includegraphics[width=.6\textwidth]{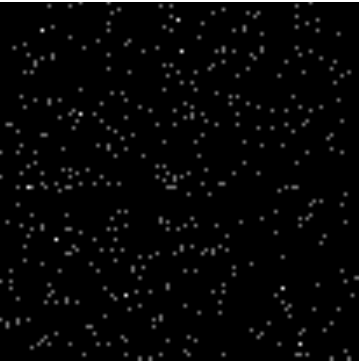}
\end{minipage}
\begin{minipage}{.49\textwidth}
\centering
\includegraphics[width=\textwidth]{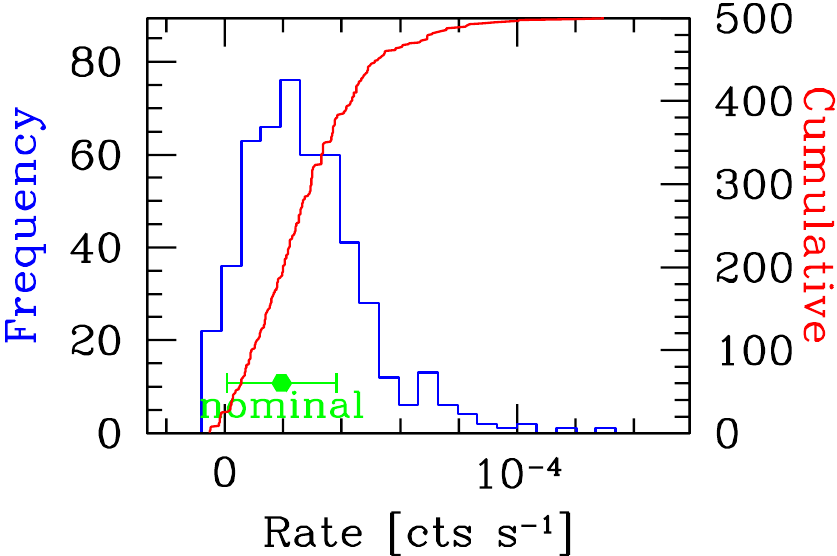}
\end{minipage}
\begin{minipage}{.49\textwidth}
\centering
\includegraphics[width=.6\textwidth]{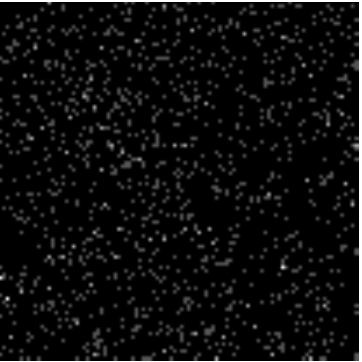}
\end{minipage}
\begin{minipage}{.49\textwidth}
\centering
\includegraphics[width=\textwidth]{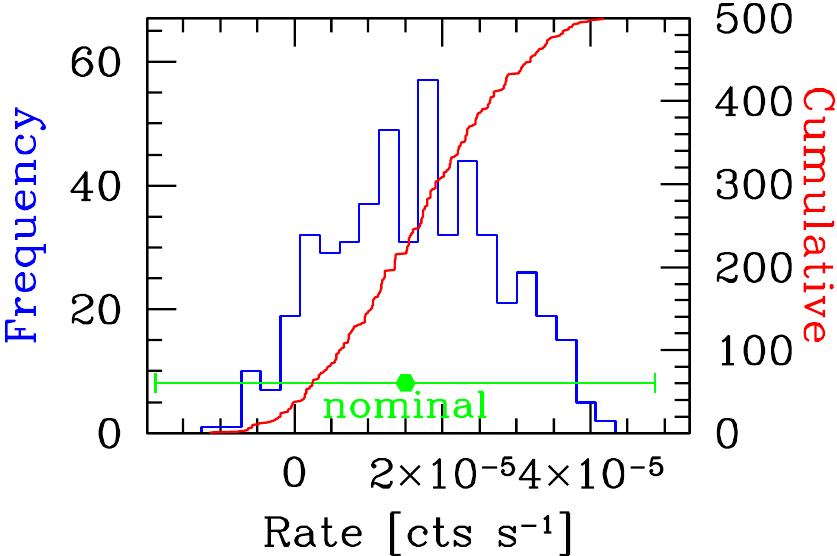}
\end{minipage}
\caption{Results of the stacking analysis performed on the {\itshape C}-COSMOS field at the coordinates of the \citet{Capak2015} high-$z$ galaxies given by \citet[][see Table \ref{table:data}]{Barisic2017}. {\itshape Upper panels:} $100''\times 100''$ stacked map of the $0.5 - 2$ keV {\itshape Chandra} passband ({\itshape left}) and corresponding count-rate histogram ({\itshape right}). {\itshape Lower panels:} the same for the $2 - 8$ keV {\itshape Chandra} passband. In both histograms, the count-rate frequencies ({\itshape blue histogram}) are shown along their cumulative distributions ({\itshape red histogram}) and their nominal mean ({\itshape green dot}) with 1$\sigma$ uncertainty.}
\label{fig:stack}
\end{figure*}

\subsection{Stacking analysis of the {\itshape Chandra} X-ray imaging}\label{sec:chandrastack}

We perform the stacking analysis of the {\itshape C}-COSMOS data at the positions of the remaining sources in order to reveal possible statistical signals, or at least get a flux upper limit that may be useful in order to plan future deep observations with X-ray facilities. For this purpose, we use the software {\scriptsize CSTACK}, a web-based tool for the stacking analysis of {\itshape Chandra} data developed by T. Miyaji\footnote{Available at {\ttfamily http://cstack.ucsd.edu/cstack/} or {\ttfamily http://lambic.astrosen.unam.mx/cstack/}.} (see Appendix \ref{sec:stack} for further details). We only stack  sources with an upper limit in $L_{\rm FIR}$, i.e., HZ1, HZ2, HZ3, HZ5a, HZ7, HZ8, and HZ8W, because  the objects with FIR detection could host heavily obscured AGN in which the X-ray emission is almost totally suppressed by the presence of high column densities of gas and IR-emitting dust \citep[e.g.,][]{Ricci2017b}. In addition, such objects are the most distant from the traditional IRX-to-$\beta$ relations adopted for galaxies, further suggesting that their spectral properties can only be explained by AGN domination.

We show the {\scriptsize CSTACK} output in Fig. \ref{fig:stack}. While no stacked signal is detected in the hard band, the soft band exhibits an average count rate of $(1.95 \pm 1.86) \cdot 10^{-5}$ cts s$^{-1}$ in the source region which is {not} compatible with zero at the $1\sigma$ level. However, due to the poor statistics of such a measurement, we are forced to treat it as an upper limit to the X-ray flux rather than a detection. We recall that at $z \sim 5.4$, the observer-frame soft band at $0.5 - 2$ keV actually observes hard $3 - 13$ keV rest-frame X-rays (for comparison, the observer-frame hard band at $2 - 8$ keV would detect photons produced in the rest-frame range $13 - 51$ keV). Adopting a mean redshift of $5.4$, an AGN-typical X-ray photon index $\Gamma \sim 1.8$ \citep[e.g.,][]{Trakhtenbrot2017} and an average Galactic $N_{\rm H} \sim 1.9 \cdot 10^{20}$ cm$^{-2}$ \citep{Dickey1990,Kalberla2005,McClure2009,Willingale2013,HI4PI2016} in the direction of the COSMOS field, the soft-band count rate upper limit of $\sim 3.8 \cdot 10^{-5}$ cts s$^{-1}$ derived from the stacking analysis corresponds to an observed flux $\lesssim 5.7 \cdot 10^{-16}$ erg s$^{-1}$ cm$^{-2}$ in the full {\itshape Chandra} band. Interpreting the soft-band excess as a detection, it corresponds to a hard X-ray luminosity of $\sim 4.6\cdot 10^{10}$ $L_{\odot}$ at $z \sim 5.4$, in excellent agreement with the luminosities reported in Table \ref{Table:Xray} for the case of AGN-dominated sources.

\section{Discussion}\label{sec:disc}

In this paper, we have presented and validated the hypothesis of star-forming galaxies with anomalous IRXs at high redshift (the zombie galaxies) hosting unveiled AGN activity. This alternative scenario to the dust temperature bias \citep{Bouwens2016} and molecule-rich dust clouds \citep{Ferrara2017} provides a natural explanation to the IRX-to-$\beta$ values of such objects in terms of variable intrinsic spectral shape and dust distribution with respect to inactive galaxies.

The comparison of the IR and UV spectral properties of the high-$z$ galaxies discovered by \citet{Capak2015} and reanalyzed by \citet{Barisic2017} with those of quasars \citep{Dong2016} and Seyferts \citep{Garcia2016} shows that high-$z$ star-forming galaxies and AGN with bolometric luminosities from $10^{43}$ to $10^{48}$ erg s$^{-1}$ populate the same region of the IRX-to-$\beta$ diagram. Furthermore, we find that the bulk of the data by \citet{Barisic2017} is described by an IRX-to-$\beta$ relation suitable for AGN-dominated objects with intrinsic spectral slopes around $\beta_{\rm int} \sim -2$ and low galactic contribution to the UV emission (i.e., $f_{\rm q} \sim 0.7$), and that this AGN UV flux is moderately reddened ($A_{1600} \lesssim 1.2$ mag). Model dependencies and degeneracies are already taken into account in the SED fitting of the AGN comparison samples, and are therefore not expected to greatly bias these conclusions (see Sect. \ref{sec:caveats}).

We further investigated the possibility of ultimately classifying the anomalous star-forming galaxies from \citet{Capak2015} as actual AGN hosts based on the correlation between their spectral properties, and on their possible X-ray emission. Though superimposing within the scatter to the AGN color-magnitude correlation derived by the \citet{Dong2016} data, the analyzed objects also lie at the boundary of the calibration region of the color-magnitude relation for inactive galaxies by \citet{Bouwens2014}. Therefore, a clear distinction between galaxy-dominated objects and AGN hosts is hard to obtain on the basis of their statistical proximity to one relation with respect to the other. We are in fact only able to classify four objects (HZ1 to HZ3 and HZ10) as normal galaxies and two (HZ5 and HZ8W) as AGN. 

Finally, since an analysis of the X-ray emission is crucial in order to definitely assess the nature of such sources, we attempted to  derive their expected X-ray luminosities in the opposite cases of AGN domination and galaxy domination, and to detect an X-ray signal from the FIR-undetected objects through stacking analysis. Interestingly, a signal not compatible with zero at the 1$\sigma$ level arises from stacking the positions of the \citet{Capak2015} objects in the {\itshape Chandra} soft band at $0.5 - 2$ keV, hinting at a possible rest-frame hard X-ray emission from at least some of the sources. According to the luminosities derived in Table \ref{Table:Xray}, such an emission is compatible with the flux expected from AGN-dominated sources at $z \sim 5.4$. Hence, future X-ray follow-ups of these objects or surveys through Athena are key to disentangling the two populations of objects.

Our work is based on the hypothesis that {all} of the zombie galaxies host an AGN at their center. Is this a overly optimistic scenario that would ultimately lead to overestimating the actual number of AGN at $z>5$? To test this scenario, we  computed the expected AGN amount in the redshift interval probed by the analyzed objects. We adopted the quasar luminosity function by \citet{Assef2011} for the interval $5 < z < 5.85$ (see, e.g., their figure 7). At those redshifts, the value of the quasar space density is  between $\sim 3.4 \cdot 10^{-6}$ and $\sim 1.1 \cdot 10^{-8}$ Mpc$^{-3}$ mag$^{-1}$ at the 1$\sigma$ level for sources with $-25 < M_J < -22.7$ (i.e., the $J$-band absolute magnitude of the zombie galaxies under the hypothesis of AGN domination), which corresponds to a number of up to 55 high-$z$ quasars in the COSMOS field of view. Restricting the area to the $C$-COSMOS value of 0.9 deg$^2$ implies a decrease of this amount to $\sim$25 objects. It should be noted that a NED search for actual quasars at $z \sim 5.4$ in the COSMOS field  resulted in two sources \citep[SDSS J095728.55+022037.8 and SDSS J100100.04+021300.6;][]{Richards2009}. Therefore, the sample of (at most) 16 high-$z$ objects discovered by \citet{Capak2015} fits inside the currently expected statistics of quasars. This topic will be further exploitable taking advantage of the upcoming releases of COSMOS spectroscopic redshifts (Salvato et al., in preparation)\footnote{We refer to  {\ttfamily http://cosmos.astro.caltech.edu/page/\\specz} for updates.}.

In addition, performing high-precision measurements with the {\itshape Very Large Baseline Interferometer} ({\itshape VLBI}) is an option to probe the radio emission of those objects, aiming at detecting possible jets associated with the accreting AGN \citep[see, e.g.,][and references therein]{Herrera2017}. In conclusion, a multiwavelength analysis of such high-redshift objects is not only desirable, but required in order to correctly estimate the star formation history and the evolution of galaxies. To this end, the {\itshape James Webb Space Telescope} \citep[{\itshape JWST};][]{Gardner2006,Kalirai2018}, the {\itshape Wide Field Infrared Survey Telescope} \citep[{\itshape WFIRST};][]{Content2013}, the {\itshape Atacama Large Millimeter/submillimeter Array} ({\itshape ALMA}), and many other facilities play a key role in acquiring new data at $z > 5$, thus opening new windows to better understand the vast, complex panorama of (mostly unknown) astronomical sources in the high-redshift Universe.

\begin{acknowledgements}
We are grateful to our anonymous referee for the helpful comments. We thank M. Ginolfi and R. Schneider (University of Rome {\it La Sapienza}) for useful discussions. FT acknowledges support from the Programma per Giovani Ricercatori -- Anno 2014 ``Rita Levi Montalcini''. This research has made use of the NASA/IPAC Extragalactic Database (NED) which is operated by the Jet Propulsion Laboratory, California Institute of Technology, under contract with the National Aeronautics and Space Administration. The development of {\scriptsize CSTACK} has been supported by CONACyT Grants No. 83564/179662, UNAM-DGAPA Grant PAPIIT IN110209/IN104113/IN104216 and {\itshape Chandra} Guest Observer Support Grant GO1-12178X. The {\itshape Chandra} X-Ray Center (CXC) is operated for NASA by the Smithsonian Astrophysical Observatory. Reproduced with permission from Astronomy \& Astrophysics, \copyright\ ESO.
\end{acknowledgements}

\bibliographystyle{aa}
\bibliography{bibliography}

\begin{thebibliography}{122}
\expandafter\ifx\csname natexlab\endcsname\relax\def\natexlab#1{#1}\fi

\bibitem[{{Abazajian} {et~al.}(2009){Abazajian}, {Adelman-McCarthy},
  {Ag{\"u}eros}, {Allam}, {Allende Prieto}, {An}, {Anderson}, {Anderson},
  {Annis}, {Bahcall}, \& et~al.}]{Abazajian2009}
{Abazajian}, K.~N., {Adelman-McCarthy}, J.~K., {Ag{\"u}eros}, M.~A., {et~al.}
  2009, \apjs, 182, 543

\bibitem[{{Adelman-McCarthy} {et~al.}(2007){Adelman-McCarthy}, {Ag{\"u}eros},
  {Allam}, {Anderson}, {Anderson}, {Annis}, {Bahcall}, {Bailer-Jones},
  {Baldry}, {Barentine}, {Beers}, {Belokurov}, {Berlind}, {Bernardi},
  {Blanton}, {Bochanski}, {Boroski}, {Bramich}, {Brewington}, {Brinchmann},
  {Brinkmann}, {Brunner}, {Budav{\'a}ri}, {Carey}, {Carliles}, {Carr},
  {Castander}, {Connolly}, {Cool}, {Cunha}, {Csabai}, {Dalcanton}, {Doi},
  {Eisenstein}, {Evans}, {Evans}, {Fan}, {Finkbeiner}, {Friedman}, {Frieman},
  {Fukugita}, {Gillespie}, {Gilmore}, {Glazebrook}, {Gray}, {Grebel}, {Gunn},
  {de Haas}, {Hall}, {Harvanek}, {Hawley}, {Hayes}, {Heckman}, {Hendry},
  {Hennessy}, {Hindsley}, {Hirata}, {Hogan}, {Hogg}, {Holtzman}, {Ichikawa},
  {Ichikawa}, {Ivezi{\'c}}, {Jester}, {Johnston}, {Jorgensen}, {Juri{\'c}},
  {Kauffmann}, {Kent}, {Kleinman}, {Knapp}, {Kniazev}, {Kron}, {Krzesinski},
  {Kuropatkin}, {Lamb}, {Lampeitl}, {Lee}, {Leger}, {Lima}, {Lin}, {Long},
  {Loveday}, {Lupton}, {Mandelbaum}, {Margon}, {Mart{\'{\i}}nez-Delgado},
  {Matsubara}, {McGehee}, {McKay}, {Meiksin}, {Munn}, {Nakajima}, {Nash},
  {Neilsen}, {Newberg}, {Nichol}, {Nieto-Santisteban}, {Nitta}, {Oyaizu},
  {Okamura}, {Ostriker}, {Padmanabhan}, {Park}, {Peoples}, {Pier}, {Pope},
  {Pourbaix}, {Quinn}, {Raddick}, {Re Fiorentin}, {Richards}, {Richmond},
  {Rix}, {Rockosi}, {Schlegel}, {Schneider}, {Scranton}, {Seljak}, {Sheldon},
  {Shimasaku}, {Silvestri}, {Smith}, {Smol{\v c}i{\'c}}, {Snedden}, {Stebbins},
  {Stoughton}, {Strauss}, {SubbaRao}, {Suto}, {Szalay}, {Szapudi}, {Szkody},
  {Tegmark}, {Thakar}, {Tremonti}, {Tucker}, {Uomoto}, {Vanden Berk},
  {Vandenberg}, {Vidrih}, {Vogeley}, {Voges}, {Vogt}, {Weinberg}, {West},
  {White}, {Wilhite}, {Yanny}, {Yocum}, {York}, {Zehavi}, {Zibetti}, \&
  {Zucker}}]{Adelman2007}
{Adelman-McCarthy}, J.~K., {Ag{\"u}eros}, M.~A., {Allam}, S.~S., {et~al.} 2007,
  \apjs, 172, 634

\bibitem[{{Ahn} {et~al.}(2014){Ahn}, {Alexandroff}, {Allende Prieto}, {Anders},
  {Anderson}, {Anderton}, {Andrews}, {Aubourg}, {Bailey}, {Bastien}, \&
  et~al.}]{Ahn2014}
{Ahn}, C.~P., {Alexandroff}, R., {Allende Prieto}, C., {et~al.} 2014, \apjs,
  211, 17

\bibitem[{{Assef} {et~al.}(2011){Assef}, {Kochanek}, {Ashby}, {Brodwin},
  {Brown}, {Cool}, {Forman}, {Gonzalez}, {Hickox}, {Jannuzi}, {Jones}, {Le
  Floc'h}, {Moustakas}, {Murray}, \& {Stern}}]{Assef2011}
{Assef}, R.~J., {Kochanek}, C.~S., {Ashby}, M.~L.~N., {et~al.} 2011, \apj, 728,
  56

\bibitem[{{Bari\v{s}i{\'c}} {et~al.}(2017){Bari\v{s}i{\'c}}, {Faisst}, {Capak},
  {Pavesi}, {Riechers}, {Scoville}, {Cooke}, {Kartaltepe}, {Casey}, \&
  {Smolcic}}]{Barisic2017}
{Bari\v{s}i{\'c}}, I., {Faisst}, A.~L., {Capak}, P.~L., {et~al.} 2017, \apj,
  845, 41

\bibitem[{{Begelman} \& {McKee}(1983)}]{Begelman1983b}
{Begelman}, M.~C. \& {McKee}, C.~F. 1983, \apj, 271, 89

\bibitem[{{Begelman} {et~al.}(1983){Begelman}, {McKee}, \&
  {Shields}}]{Begelman1983a}
{Begelman}, M.~C., {McKee}, C.~F., \& {Shields}, G.~A. 1983, \apj, 271, 70

\bibitem[{{Behrens} {et~al.}(2018){Behrens}, {Pallottini}, {Ferrara},
  {Gallerani}, \& {Vallini}}]{Behrens2018}
{Behrens}, C., {Pallottini}, A., {Ferrara}, A., {Gallerani}, S., \& {Vallini},
  L. 2018, \mnras, 477, 552

\bibitem[{{Bianchi} \& {Schneider}(2007)}]{Bianchi2007}
{Bianchi}, S. \& {Schneider}, R. 2007, \mnras, 378, 973

\bibitem[{{Bouwens} {et~al.}(2016){Bouwens}, {Aravena}, {Decarli}, {Walter},
  {da Cunha}, {Labb{\'e}}, {Bauer}, {Bertoldi}, {Carilli}, {Chapman}, {Daddi},
  {Hodge}, {Ivison}, {Karim}, {Le Fevre}, {Magnelli}, {Ota}, {Riechers},
  {Smail}, {van der Werf}, {Weiss}, {Cox}, {Elbaz}, {Gonzalez-Lopez},
  {Infante}, {Oesch}, {Wagg}, \& {Wilkins}}]{Bouwens2016}
{Bouwens}, R.~J., {Aravena}, M., {Decarli}, R., {et~al.} 2016, \apj, 833, 72

\bibitem[{{Bouwens} {et~al.}(2014){Bouwens}, {Illingworth}, {Oesch},
  {Labb{\'e}}, {van Dokkum}, {Trenti}, {Franx}, {Smit}, {Gonzalez}, \&
  {Magee}}]{Bouwens2014}
{Bouwens}, R.~J., {Illingworth}, G.~D., {Oesch}, P.~A., {et~al.} 2014, \apj,
  793, 115

\bibitem[{{Bouwens} {et~al.}(2015){Bouwens}, {Illingworth}, {Oesch}, {Trenti},
  {Labb{\'e}}, {Bradley}, {Carollo}, {van Dokkum}, {Gonzalez}, {Holwerda},
  {Franx}, {Spitler}, {Smit}, \& {Magee}}]{Bouwens2015a}
{Bouwens}, R.~J., {Illingworth}, G.~D., {Oesch}, P.~A., {et~al.} 2015, \apj,
  803, 34

\bibitem[{{Bowler} {et~al.}(2015){Bowler}, {Dunlop}, {McLure}, {McCracken},
  {Milvang-Jensen}, {Furusawa}, {Taniguchi}, {Le F{\`e}vre}, {Fynbo}, {Jarvis},
  \& {H{\"a}u{\ss}ler}}]{Bowler2015}
{Bowler}, R.~A.~A., {Dunlop}, J.~S., {McLure}, R.~J., {et~al.} 2015, \mnras,
  452, 1817

\bibitem[{{Brightman} {et~al.}(2014){Brightman}, {Nandra}, {Salvato}, {Hsu},
  {Aird}, \& {Rangel}}]{Brightman2014}
{Brightman}, M., {Nandra}, K., {Salvato}, M., {et~al.} 2014, \mnras, 443, 1999

\bibitem[{{Calzetti}(1997)}]{Calzetti1997}
{Calzetti}, D. 1997, \aj, 113, 162

\bibitem[{{Capak} {et~al.}(2015){Capak}, {Carilli}, {Jones}, {Casey},
  {Riechers}, {Sheth}, {Carollo}, {Ilbert}, {Karim}, {Lefevre}, {Lilly},
  {Scoville}, {Smolcic}, \& {Yan}}]{Capak2015}
{Capak}, P.~L., {Carilli}, C., {Jones}, G., {et~al.} 2015, \nat, 522, 455

\bibitem[{{Casey} {et~al.}(2014){Casey}, {Scoville}, {Sanders}, {Lee},
  {Cooray}, {Finkelstein}, {Capak}, {Conley}, {De Zotti}, {Farrah}, {Fu}, {Le
  Floc'h}, {Ilbert}, {Ivison}, \& {Takeuchi}}]{Casey2014}
{Casey}, C.~M., {Scoville}, N.~Z., {Sanders}, D.~B., {et~al.} 2014, \apj, 796,
  95

\bibitem[{{Cattaneo} {et~al.}(2009){Cattaneo}, {Faber}, {Binney}, {Dekel},
  {Kormendy}, {Mushotzky}, {Babul}, {Best}, {Br{\"u}ggen}, {Fabian}, {Frenk},
  {Khalatyan}, {Netzer}, {Mahdavi}, {Silk}, {Steinmetz}, \&
  {Wisotzki}}]{Cattaneo2009}
{Cattaneo}, A., {Faber}, S.~M., {Binney}, J., {et~al.} 2009, \nat, 460, 213

\bibitem[{{Civano} {et~al.}(2012){Civano}, {Elvis}, {Brusa}, {Comastri},
  {Salvato}, {Zamorani}, {Aldcroft}, {Bongiorno}, {Capak}, {Cappelluti},
  {Cisternas}, {Fiore}, {Fruscione}, {Hao}, {Kartaltepe}, {Koekemoer}, {Gilli},
  {Impey}, {Lanzuisi}, {Lusso}, {Mainieri}, {Miyaji}, {Lilly}, {Masters},
  {Puccetti}, {Schawinski}, {Scoville}, {Silverman}, {Trump}, {Urry},
  {Vignali}, \& {Wright}}]{Civano2012}
{Civano}, F., {Elvis}, M., {Brusa}, M., {et~al.} 2012, \apjs, 201, 30

\bibitem[{{Content} {et~al.}(2013){Content}, {Aaron}, {Abplanalp}, {Anderson},
  {Capps}, {Chang}, {Dooley}, {Egerman}, {Goullioud}, {Klein}, {Kruk}, {Kuan},
  {Melton}, {Ruffa}, {Underhill}, \& {Van Buren}}]{Content2013}
{Content}, D., {Aaron}, K., {Abplanalp}, L., {et~al.} 2013, in \procspie, Vol.
  8860, UV/Optical/IR Space Telescopes and Instruments: Innovative Technologies
  and Concepts VI, 88600E

\bibitem[{{Cullen} {et~al.}(2017){Cullen}, {McLure}, {Khochfar}, {Dunlop}, \&
  {Dalla Vecchia}}]{Cullen2017}
{Cullen}, F., {McLure}, R.~J., {Khochfar}, S., {Dunlop}, J.~S., \& {Dalla
  Vecchia}, C. 2017, \mnras, 470, 3006

\bibitem[{{Dickey} \& {Lockman}(1990)}]{Dickey1990}
{Dickey}, J.~M. \& {Lockman}, F.~J. 1990, \araa, 28, 215

\bibitem[{{Dong} \& {Wu}(2016)}]{Dong2016}
{Dong}, X.~Y. \& {Wu}, X.-B. 2016, \apj, 824, 70

\bibitem[{{Draine}(2011)}]{draine2011}
{Draine}, B.~T. 2011, Physics of the Interstellar and Intergalactic Medium
  (Princeton Univ. Press)

\bibitem[{{Duras} {et~al.}(2017){Duras}, {Bongiorno}, {Piconcelli}, {Bianchi},
  {Pappalardo}, {Valiante}, {Bischetti}, {Feruglio}, {Martocchia}, {Schneider},
  {Vietri}, {Vignali}, {Zappacosta}, {La Franca}, \& {Fiore}}]{Duras2017}
{Duras}, F., {Bongiorno}, A., {Piconcelli}, E., {et~al.} 2017, \aap, 604, A67

\bibitem[{{Elvis} {et~al.}(2009){Elvis}, {Civano}, {Vignali}, {Puccetti},
  {Fiore}, {Cappelluti}, {Aldcroft}, {Fruscione}, {Zamorani}, {Comastri},
  {Brusa}, {Gilli}, {Miyaji}, {Damiani}, {Koekemoer}, {Finoguenov}, {Brunner},
  {Urry}, {Silverman}, {Mainieri}, {Hasinger}, {Griffiths}, {Carollo}, {Hao},
  {Guzzo}, {Blain}, {Calzetti}, {Carilli}, {Capak}, {Ettori}, {Fabbiano},
  {Impey}, {Lilly}, {Mobasher}, {Rich}, {Salvato}, {Sanders}, {Schinnerer},
  {Scoville}, {Shopbell}, {Taylor}, {Taniguchi}, \& {Volonteri}}]{Elvis2009}
{Elvis}, M., {Civano}, F., {Vignali}, C., {et~al.} 2009, \apjs, 184, 158

\bibitem[{{Fabian}(2012)}]{Fabian2012}
{Fabian}, A.~C. 2012, \araa, 50, 455

\bibitem[{{Ferrara} {et~al.}(2017){Ferrara}, {Hirashita}, {Ouchi}, \&
  {Fujimoto}}]{Ferrara2017}
{Ferrara}, A., {Hirashita}, H., {Ouchi}, M., \& {Fujimoto}, S. 2017, \mnras,
  471, 5018

\bibitem[{{Finkelstein} {et~al.}(2015{\natexlab{a}}){Finkelstein}, {Ryan},
  {Papovich}, {Dickinson}, {Song}, {Somerville}, {Ferguson}, {Salmon},
  {Giavalisco}, {Koekemoer}, {Ashby}, {Behroozi}, {Castellano}, {Dunlop},
  {Faber}, {Fazio}, {Fontana}, {Grogin}, {Hathi}, {Jaacks}, {Kocevski},
  {Livermore}, {McLure}, {Merlin}, {Mobasher}, {Newman}, {Rafelski}, {Tilvi},
  \& {Willner}}]{Finkelstein2015}
{Finkelstein}, S.~L., {Ryan}, Jr., R.~E., {Papovich}, C., {et~al.}
  2015{\natexlab{a}}, \apj, 810, 71

\bibitem[{{Finkelstein} {et~al.}(2015{\natexlab{b}}){Finkelstein}, {Song},
  {Behroozi}, {Somerville}, {Papovich}, {Milosavljevi{\'c}}, {Dekel},
  {Narayanan}, {Ashby}, {Cooray}, {Fazio}, {Ferguson}, {Koekemoer}, {Salmon},
  \& {Willner}}]{Finkelstein2015b}
{Finkelstein}, S.~L., {Song}, M., {Behroozi}, P., {et~al.} 2015{\natexlab{b}},
  \apj, 814, 95

\bibitem[{{Forrest} {et~al.}(2016){Forrest}, {Tran}, {Tomczak}, {Broussard},
  {Labb{\'e}}, {Papovich}, {Kriek}, {Allen}, {Cowley}, {Dickinson},
  {Glazebrook}, {van Houdt}, {Inami}, {Kacprzak}, {Kawinwanichakij}, {Kelson},
  {McCarthy}, {Monson}, {Morrison}, {Nanayakkara}, {Persson}, {Quadri},
  {Spitler}, {Straatman}, \& {Tilvi}}]{Forrest2016}
{Forrest}, B., {Tran}, K.-V.~H., {Tomczak}, A.~R., {et~al.} 2016, \apjl, 818,
  L26

\bibitem[{{Gallerani} {et~al.}(2010){Gallerani}, {Maiolino}, {Juarez}, {Nagao},
  {Marconi}, {Bianchi}, {Schneider}, {Mannucci}, {Oliva}, {Willott}, {Jiang},
  \& {Fan}}]{Gallerani2010}
{Gallerani}, S., {Maiolino}, R., {Juarez}, Y., {et~al.} 2010, \aap, 523, A85

\bibitem[{{Garc{\'{\i}}a-Gonz{\'a}lez}
  {et~al.}(2016){Garc{\'{\i}}a-Gonz{\'a}lez}, {Alonso-Herrero},
  {Hern{\'a}n-Caballero}, {Pereira-Santaella}, {Ramos-Almeida},
  {Acosta-Pulido}, {D{\'{\i}}az-Santos}, {Esquej},
  {Gonz{\'a}lez-Mart{\'{\i}}n}, {Ichikawa}, {L{\'o}pez-Rodr{\'{\i}}guez},
  {Povic}, {Roche}, \& {S{\'a}nchez-Portal}}]{Garcia2016}
{Garc{\'{\i}}a-Gonz{\'a}lez}, J., {Alonso-Herrero}, A., {Hern{\'a}n-Caballero},
  A., {et~al.} 2016, \mnras, 458, 4512

\bibitem[{{Gardner} {et~al.}(2006){Gardner}, {Mather}, {Clampin}, {Doyon},
  {Greenhouse}, {Hammel}, {Hutchings}, {Jakobsen}, {Lilly}, {Long}, {Lunine},
  {McCaughrean}, {Mountain}, {Nella}, {Rieke}, {Rieke}, {Rix}, {Smith},
  {Sonneborn}, {Stiavelli}, {Stockman}, {Windhorst}, \& {Wright}}]{Gardner2006}
{Gardner}, J.~P., {Mather}, J.~C., {Clampin}, M., {et~al.} 2006, \ssr, 123, 485

\bibitem[{{Gil de Paz} {et~al.}(2007){Gil de Paz}, {Boissier}, {Madore},
  {Seibert}, {Joe}, {Boselli}, {Wyder}, {Thilker}, {Bianchi}, {Rey}, {Rich},
  {Barlow}, {Conrow}, {Forster}, {Friedman}, {Martin}, {Morrissey}, {Neff},
  {Schiminovich}, {Small}, {Donas}, {Heckman}, {Lee}, {Milliard}, {Szalay}, \&
  {Yi}}]{Gildepaz2007}
{Gil de Paz}, A., {Boissier}, S., {Madore}, B.~F., {et~al.} 2007, \apjs, 173,
  185

\bibitem[{{Gilli} {et~al.}(2007){Gilli}, {Comastri}, \& {Hasinger}}]{Gilli2007}
{Gilli}, R., {Comastri}, A., \& {Hasinger}, G. 2007, \aap, 463, 79

\bibitem[{{Giustini} {et~al.}(2008){Giustini}, {Cappi}, \&
  {Vignali}}]{Giustini2008}
{Giustini}, M., {Cappi}, M., \& {Vignali}, C. 2008, \aap, 491, 425

\bibitem[{{Glikman} {et~al.}(2006){Glikman}, {Helfand}, \&
  {White}}]{Glikman2006}
{Glikman}, E., {Helfand}, D.~J., \& {White}, R.~L. 2006, \apj, 640, 579

\bibitem[{{Gordon} {et~al.}(1997){Gordon}, {Calzetti}, \& {Witt}}]{Gordon1997}
{Gordon}, K.~D., {Calzetti}, D., \& {Witt}, A.~N. 1997, \apj, 487, 625

\bibitem[{{Halpern}(1982)}]{Halpern1982}
{Halpern}, J.~P. 1982, PhD thesis, Harvard University, Cambridge, MA.

\bibitem[{{Hao} {et~al.}(2013){Hao}, {Elvis}, {Bongiorno}, {Zamorani},
  {Merloni}, {Kelly}, {Civano}, {Celotti}, {Ho}, {Jahnke}, {Comastri}, {Trump},
  {Mainieri}, {Salvato}, {Brusa}, {Impey}, {Koekemoer}, {Lanzuisi}, {Vignali},
  {Silverman}, {Urry}, \& {Schawinski}}]{Hao2013}
{Hao}, H., {Elvis}, M., {Bongiorno}, A., {et~al.} 2013, \mnras, 434, 3104

\bibitem[{{Herrera Ruiz} {et~al.}(2017){Herrera Ruiz}, {Middelberg}, {Deller},
  {Norris}, {Best}, {Brisken}, {Schinnerer}, {Smol{\v c}i{\'c}}, {Delvecchio},
  {Momjian}, {Bomans}, {Scoville}, \& {Carilli}}]{Herrera2017}
{Herrera Ruiz}, N., {Middelberg}, E., {Deller}, A., {et~al.} 2017, \aap, 607,
  A132

\bibitem[{{HI4PI Collaboration} {et~al.}(2016){HI4PI Collaboration}, {Ben
  Bekhti}, {Fl{\"o}er}, {Keller}, {Kerp}, {Lenz}, {Winkel}, {Bailin},
  {Calabretta}, {Dedes}, {Ford}, {Gibson}, {Haud}, {Janowiecki}, {Kalberla},
  {Lockman}, {McClure-Griffiths}, {Murphy}, {Nakanishi}, {Pisano}, \&
  {Staveley-Smith}}]{HI4PI2016}
{HI4PI Collaboration}, {Ben Bekhti}, N., {Fl{\"o}er}, L., {et~al.} 2016, \aap,
  594, A116

\bibitem[{{Inoue}(2005)}]{Inoue2005}
{Inoue}, A.~K. 2005, \mnras, 359, 171

\bibitem[{{Ishibashi} {et~al.}(2017){Ishibashi}, {Banerji}, \&
  {Fabian}}]{Ishibashi2017}
{Ishibashi}, W., {Banerji}, M., \& {Fabian}, A.~C. 2017, \mnras, 469, 1496

\bibitem[{{Ishibashi} \& {Fabian}(2016)}]{Ishibashi2016}
{Ishibashi}, W. \& {Fabian}, A.~C. 2016, \mnras, 457, 2864

\bibitem[{{Kalberla} {et~al.}(2005){Kalberla}, {Burton}, {Hartmann}, {Arnal},
  {Bajaja}, {Morras}, \& {P{\"o}ppel}}]{Kalberla2005}
{Kalberla}, P.~M.~W., {Burton}, W.~B., {Hartmann}, D., {et~al.} 2005, \aap,
  440, 775

\bibitem[{{Kalfountzou} {et~al.}(2014){Kalfountzou}, {Civano}, {Elvis},
  {Trichas}, \& {Green}}]{Kalfountzou2014}
{Kalfountzou}, E., {Civano}, F., {Elvis}, M., {Trichas}, M., \& {Green}, P.
  2014, \mnras, 445, 1430

\bibitem[{{Kalirai}(2018)}]{Kalirai2018}
{Kalirai}, J. 2018, ArXiv e-prints [\eprint[arXiv]{1805.06941}]

\bibitem[{{Kennicutt} \& {Evans}(2012)}]{Kennicutt2012}
{Kennicutt}, R.~C. \& {Evans}, N.~J. 2012, \araa, 50, 531

\bibitem[{{Knudsen} {et~al.}(2017){Knudsen}, {Watson}, {Frayer}, {Christensen},
  {Gallazzi}, {Micha{\l}owski}, {Richard}, \& {Zavala}}]{Knudsen2017}
{Knudsen}, K.~K., {Watson}, D., {Frayer}, D., {et~al.} 2017, \mnras, 466, 138

\bibitem[{{Krumpe} {et~al.}(2014){Krumpe}, {Markowitz}, \&
  {Nikutta}}]{Krumpe2014}
{Krumpe}, M., {Markowitz}, A., \& {Nikutta}, R. 2014, in IAU Symposium, Vol.
  304, Multiwavelength AGN Surveys and Studies, ed. A.~M. {Mickaelian} \& D.~B.
  {Sanders}, 265--265

\bibitem[{{Lanzuisi} {et~al.}(2013){Lanzuisi}, {Civano}, {Elvis}, {Salvato},
  {Hasinger}, {Vignali}, {Zamorani}, {Aldcroft}, {Brusa}, {Comastri}, {Fiore},
  {Fruscione}, {Gilli}, {Ho}, {Mainieri}, {Merloni}, \&
  {Siemiginowska}}]{Lanzuisi2013}
{Lanzuisi}, G., {Civano}, F., {Elvis}, M., {et~al.} 2013, \mnras, 431, 978

\bibitem[{{Lawrence} {et~al.}(2007){Lawrence}, {Warren}, {Almaini}, {Edge},
  {Hambly}, {Jameson}, {Lucas}, {Casali}, {Adamson}, {Dye}, {Emerson},
  {Foucaud}, {Hewett}, {Hirst}, {Hodgkin}, {Irwin}, {Lodieu}, {McMahon},
  {Simpson}, {Smail}, {Mortlock}, \& {Folger}}]{Lawrence2007}
{Lawrence}, A., {Warren}, S.~J., {Almaini}, O., {et~al.} 2007, \mnras, 379,
  1599

\bibitem[{{Lehmer} {et~al.}(2010){Lehmer}, {Alexander}, {Bauer}, {Brandt},
  {Goulding}, {Jenkins}, {Ptak}, \& {Roberts}}]{Lehmer2010}
{Lehmer}, B.~D., {Alexander}, D.~M., {Bauer}, F.~E., {et~al.} 2010, \apj, 724,
  559

\bibitem[{{Lehmer} {et~al.}(2016){Lehmer}, {Basu-Zych}, {Mineo}, {Brandt},
  {Eufrasio}, {Fragos}, {Hornschemeier}, {Luo}, {Xue}, {Bauer}, {Gilfanov},
  {Ranalli}, {Schneider}, {Shemmer}, {Tozzi}, {Trump}, {Vignali}, {Wang},
  {Yukita}, \& {Zezas}}]{Lehmer2016}
{Lehmer}, B.~D., {Basu-Zych}, A.~R., {Mineo}, S., {et~al.} 2016, \apj, 825, 7

\bibitem[{{Lusso} {et~al.}(2010){Lusso}, {Comastri}, {Vignali}, {Zamorani},
  {Brusa}, {Gilli}, {Iwasawa}, {Salvato}, {Civano}, {Elvis}, {Merloni},
  {Bongiorno}, {Trump}, {Koekemoer}, {Schinnerer}, {Le Floc'h}, {Cappelluti},
  {Jahnke}, {Sargent}, {Silverman}, {Mainieri}, {Fiore}, {Bolzonella}, {Le
  F{\`e}vre}, {Garilli}, {Iovino}, {Kneib}, {Lamareille}, {Lilly}, {Mignoli},
  {Scodeggio}, \& {Vergani}}]{Lusso2010}
{Lusso}, E., {Comastri}, A., {Vignali}, C., {et~al.} 2010, \aap, 512, A34

\bibitem[{{Lusso} \& {Risaliti}(2016)}]{Lusso2016}
{Lusso}, E. \& {Risaliti}, G. 2016, \apj, 819, 154

\bibitem[{{Ma} {et~al.}(2016){Ma}, {Hopkins}, {Faucher-Gigu{\`e}re}, {Zolman},
  {Muratov}, {Kere{\v s}}, \& {Quataert}}]{Ma2016}
{Ma}, X., {Hopkins}, P.~F., {Faucher-Gigu{\`e}re}, C.-A., {et~al.} 2016,
  \mnras, 456, 2140

\bibitem[{{Mainzer} {et~al.}(2011){Mainzer}, {Bauer}, {Grav}, {Masiero},
  {Cutri}, {Dailey}, {Eisenhardt}, {McMillan}, {Wright}, {Walker}, {Jedicke},
  {Spahr}, {Tholen}, {Alles}, {Beck}, {Brandenburg}, {Conrow}, {Evans},
  {Fowler}, {Jarrett}, {Marsh}, {Masci}, {McCallon}, {Wheelock}, {Wittman},
  {Wyatt}, {DeBaun}, {Elliott}, {Elsbury}, {Gautier}, {Gomillion}, {Leisawitz},
  {Maleszewski}, {Micheli}, \& {Wilkins}}]{Mainzer2011}
{Mainzer}, A., {Bauer}, J., {Grav}, T., {et~al.} 2011, \apj, 731, 53

\bibitem[{{Maiolino} \& {Rieke}(1995)}]{Maiolino1995}
{Maiolino}, R. \& {Rieke}, G.~H. 1995, \apj, 454, 95

\bibitem[{{Mancini} {et~al.}(2016){Mancini}, {Schneider}, {Graziani},
  {Valiante}, {Dayal}, {Maio}, \& {Ciardi}}]{Mancini2016}
{Mancini}, M., {Schneider}, R., {Graziani}, L., {et~al.} 2016, \mnras, 462,
  3130

\bibitem[{{Mancini} {et~al.}(2015){Mancini}, {Schneider}, {Graziani},
  {Valiante}, {Dayal}, {Maio}, {Ciardi}, \& {Hunt}}]{mancini2015}
{Mancini}, M., {Schneider}, R., {Graziani}, L., {et~al.} 2015, \mnras, 451, L70

\bibitem[{{Mannucci} {et~al.}(2010){Mannucci}, {Cresci}, {Maiolino}, {Marconi},
  \& {Gnerucci}}]{Mannucci2010}
{Mannucci}, F., {Cresci}, G., {Maiolino}, R., {Marconi}, A., \& {Gnerucci}, A.
  2010, \mnras, 408, 2115

\bibitem[{{Marchesi} {et~al.}(2016){Marchesi}, {Civano}, {Elvis}, {Salvato},
  {Brusa}, {Comastri}, {Gilli}, {Hasinger}, {Lanzuisi}, {Miyaji}, {Treister},
  {Urry}, {Vignali}, {Zamorani}, {Allevato}, {Cappelluti}, {Cardamone},
  {Finoguenov}, {Griffiths}, {Karim}, {Laigle}, {LaMassa}, {Jahnke}, {Ranalli},
  {Schawinski}, {Schinnerer}, {Silverman}, {Smolcic}, {Suh}, \&
  {Trakhtenbrot}}]{Marchesi2016}
{Marchesi}, S., {Civano}, F., {Elvis}, M., {et~al.} 2016, \apj, 817, 34

\bibitem[{{Marconi} {et~al.}(2004){Marconi}, {Risaliti}, {Gilli}, {Hunt},
  {Maiolino}, \& {Salvati}}]{Marconi2004}
{Marconi}, A., {Risaliti}, G., {Gilli}, R., {et~al.} 2004, \mnras, 351, 169

\bibitem[{{Martin} {et~al.}(2005){Martin}, {Fanson}, {Schiminovich},
  {Morrissey}, {Friedman}, {Barlow}, {Conrow}, {Grange}, {Jelinsky},
  {Milliard}, {Siegmund}, {Bianchi}, {Byun}, {Donas}, {Forster}, {Heckman},
  {Lee}, {Madore}, {Malina}, {Neff}, {Rich}, {Small}, {Surber}, {Szalay},
  {Welsh}, \& {Wyder}}]{Martin2005}
{Martin}, D.~C., {Fanson}, J., {Schiminovich}, D., {et~al.} 2005, \apjl, 619,
  L1

\bibitem[{{McClure-Griffiths} {et~al.}(2009){McClure-Griffiths}, {Pisano},
  {Calabretta}, {Ford}, {Lockman}, {Staveley-Smith}, {Kalberla}, {Bailin},
  {Dedes}, {Janowiecki}, {Gibson}, {Murphy}, {Nakanishi}, \&
  {Newton-McGee}}]{McClure2009}
{McClure-Griffiths}, N.~M., {Pisano}, D.~J., {Calabretta}, M.~R., {et~al.}
  2009, \apjs, 181, 398

\bibitem[{{Meurer} {et~al.}(1999){Meurer}, {Heckman}, \&
  {Calzetti}}]{Meurer1999}
{Meurer}, G.~R., {Heckman}, T.~M., \& {Calzetti}, D. 1999, \apj, 521, 64

\bibitem[{{Morrison} \& {McCammon}(1983)}]{Morrison1983}
{Morrison}, R. \& {McCammon}, D. 1983, \apj, 270, 119

\bibitem[{{Mullaney} {et~al.}(2011){Mullaney}, {Alexander}, {Goulding}, \&
  {Hickox}}]{Mullaney2011}
{Mullaney}, J.~R., {Alexander}, D.~M., {Goulding}, A.~D., \& {Hickox}, R.~C.
  2011, \mnras, 414, 1082

\bibitem[{{Narayanan} {et~al.}(2018){Narayanan}, {Dav{\'e}}, {Johnson},
  {Thompson}, {Conroy}, \& {Geach}}]{Narayanan2017}
{Narayanan}, D., {Dav{\'e}}, R., {Johnson}, B.~D., {et~al.} 2018, \mnras, 474,
  1718

\bibitem[{{Ouchi} {et~al.}(1999){Ouchi}, {Yamada}, {Kawai}, \&
  {Ohta}}]{Ouchi1999}
{Ouchi}, M., {Yamada}, T., {Kawai}, H., \& {Ohta}, K. 1999, \apjl, 517, L19

\bibitem[{{Page} {et~al.}(2017){Page}, {Carrera}, {Ceballos}, {Corral},
  {Ebrero}, {Esquej}, {Krumpe}, {Mateos}, {Rosen}, {Schwope}, {Streblyanska},
  {Symeonidis}, {Tedds}, \& {Watson}}]{Page2017}
{Page}, M.~J., {Carrera}, F.~J., {Ceballos}, M., {et~al.} 2017, \mnras, 464,
  4586

\bibitem[{{P{\^a}ris} {et~al.}(2014){P{\^a}ris}, {Petitjean}, {Aubourg},
  {Ross}, {Myers}, {Streblyanska}, {Bailey}, {Hall}, {Strauss}, {Anderson},
  {Bizyaev}, {Borde}, {Brinkmann}, {Bovy}, {Brandt}, {Brewington},
  {Brownstein}, {Cook}, {Ebelke}, {Fan}, {Filiz Ak}, {Finley}, {Font-Ribera},
  {Ge}, {Hamann}, {Ho}, {Jiang}, {Kinemuchi}, {Malanushenko}, {Malanushenko},
  {Marchante}, {McGreer}, {McMahon}, {Miralda-Escud{\'e}}, {Muna},
  {Noterdaeme}, {Oravetz}, {Palanque-Delabrouille}, {Pan}, {Perez-Fournon},
  {Pieri}, {Riffel}, {Schlegel}, {Schneider}, {Simmons}, {Viel}, {Weaver},
  {Wood-Vasey}, {Y{\`e}che}, \& {York}}]{Paris2014}
{P{\^a}ris}, I., {Petitjean}, P., {Aubourg}, {\'E}., {et~al.} 2014, \aap, 563,
  A54

\bibitem[{{Pei}(1992)}]{Pei1992}
{Pei}, Y.~C. 1992, \apj, 395, 130

\bibitem[{{Penner} {et~al.}(2012){Penner}, {Dickinson}, {Pope}, {Dey},
  {Magnelli}, {Pannella}, {Altieri}, {Aussel}, {Buat}, {Bussmann},
  {Charmandaris}, {Coia}, {Daddi}, {Dannerbauer}, {Elbaz}, {Hwang},
  {Kartaltepe}, {Lin}, {Magdis}, {Morrison}, {Popesso}, {Scott}, \&
  {Valtchanov}}]{Penner2012}
{Penner}, K., {Dickinson}, M., {Pope}, A., {et~al.} 2012, \apj, 759, 28

\bibitem[{{Pettini} {et~al.}(1998){Pettini}, {Kellogg}, {Steidel}, {Dickinson},
  {Adelberger}, \& {Giavalisco}}]{Pettini1998}
{Pettini}, M., {Kellogg}, M., {Steidel}, C.~C., {et~al.} 1998, \apj, 508, 539

\bibitem[{{Pezzulli} {et~al.}(2017){Pezzulli}, {Valiante}, {Orofino},
  {Schneider}, {Gallerani}, \& {Sbarrato}}]{Pezzulli2017}
{Pezzulli}, E., {Valiante}, R., {Orofino}, M.~C., {et~al.} 2017, \mnras, 466,
  2131

\bibitem[{{Popping} {et~al.}(2017){Popping}, {Puglisi}, \&
  {Norman}}]{Popping2017}
{Popping}, G., {Puglisi}, A., \& {Norman}, C.~A. 2017, \mnras, 472, 2315

\bibitem[{{Puccetti} {et~al.}(2009){Puccetti}, {Vignali}, {Cappelluti},
  {Fiore}, {Zamorani}, {Aldcroft}, {Elvis}, {Gilli}, {Miyaji}, {Brunner},
  {Brusa}, {Civano}, {Comastri}, {Damiani}, {Fruscione}, {Finoguenov},
  {Koekemoer}, \& {Mainieri}}]{Puccetti2009}
{Puccetti}, S., {Vignali}, C., {Cappelluti}, N., {et~al.} 2009, \apjs, 185, 586

\bibitem[{{Ranalli} {et~al.}(2003){Ranalli}, {Comastri}, \&
  {Setti}}]{Ranalli2003}
{Ranalli}, P., {Comastri}, A., \& {Setti}, G. 2003, \aap, 399, 39

\bibitem[{{Ranalli} {et~al.}(2005){Ranalli}, {Comastri}, \&
  {Setti}}]{Ranalli2005}
{Ranalli}, P., {Comastri}, A., \& {Setti}, G. 2005, \aap, 440, 23

\bibitem[{{Reddy} {et~al.}(2015){Reddy}, {Kriek}, {Shapley}, {Freeman},
  {Siana}, {Coil}, {Mobasher}, {Price}, {Sanders}, \& {Shivaei}}]{Reddy2015}
{Reddy}, N.~A., {Kriek}, M., {Shapley}, A.~E., {et~al.} 2015, \apj, 806, 259

\bibitem[{{Reddy} {et~al.}(2018){Reddy}, {Oesch}, {Bouwens}, {Montes},
  {Illingworth}, {Steidel}, {van Dokkum}, {Atek}, {Carollo}, {Cibinel},
  {Holden}, {Labb{\'e}}, {Magee}, {Morselli}, {Nelson}, \&
  {Wilkins}}]{Reddy2017}
{Reddy}, N.~A., {Oesch}, P.~A., {Bouwens}, R.~J., {et~al.} 2018, \apj, 853, 56

\bibitem[{{Ricci} {et~al.}(2017{\natexlab{a}}){Ricci}, {Bauer}, {Treister},
  {Schawinski}, {Privon}, {Blecha}, {Arevalo}, {Armus}, {Harrison}, {Ho},
  {Iwasawa}, {Sanders}, \& {Stern}}]{Ricci2017b}
{Ricci}, C., {Bauer}, F.~E., {Treister}, E., {et~al.} 2017{\natexlab{a}},
  \mnras, 468, 1273

\bibitem[{{Ricci} {et~al.}(2017{\natexlab{b}}){Ricci}, {Trakhtenbrot}, {Koss},
  {Ueda}, {Schawinski}, {Oh}, {Lamperti}, {Mushotzky}, {Treister}, {Ho},
  {Weigel}, {Bauer}, {Paltani}, {Fabian}, {Xie}, \& {Gehrels}}]{Ricci2017a}
{Ricci}, C., {Trakhtenbrot}, B., {Koss}, M.~J., {et~al.} 2017{\natexlab{b}},
  \nat, 549, 488

\bibitem[{{Richards} {et~al.}(2009){Richards}, {Deo}, {Lacy}, {Myers},
  {Nichol}, {Zakamska}, {Brunner}, {Brandt}, {Gray}, {Parejko}, {Ptak},
  {Schneider}, {Storrie-Lombardi}, \& {Szalay}}]{Richards2009}
{Richards}, G.~T., {Deo}, R.~P., {Lacy}, M., {et~al.} 2009, \aj, 137, 3884

\bibitem[{{Risaliti} {et~al.}(2011){Risaliti}, {Salvati}, \&
  {Marconi}}]{Risaliti2011}
{Risaliti}, G., {Salvati}, M., \& {Marconi}, A. 2011, \mnras, 411, 2223

\bibitem[{{Rogers} {et~al.}(2014){Rogers}, {McLure}, {Dunlop}, {Bowler},
  {Curtis-Lake}, {Dayal}, {Faber}, {Ferguson}, {Finkelstein}, {Grogin},
  {Hathi}, {Kocevski}, {Koekemoer}, \& {Kurczynski}}]{Rogers2014}
{Rogers}, A.~B., {McLure}, R.~J., {Dunlop}, J.~S., {et~al.} 2014, \mnras, 440,
  3714

\bibitem[{{Runnoe} {et~al.}(2012{\natexlab{a}}){Runnoe}, {Brotherton}, \&
  {Shang}}]{Runnoe2012c}
{Runnoe}, J.~C., {Brotherton}, M.~S., \& {Shang}, Z. 2012{\natexlab{a}},
  \mnras, 427, 1800

\bibitem[{{Runnoe} {et~al.}(2012{\natexlab{b}}){Runnoe}, {Brotherton}, \&
  {Shang}}]{Runnoe2012}
{Runnoe}, J.~C., {Brotherton}, M.~S., \& {Shang}, Z. 2012{\natexlab{b}},
  \mnras, 422, 478

\bibitem[{{Salpeter}(1964)}]{Salpeter1964}
{Salpeter}, E.~E. 1964, \apj, 140, 796

\bibitem[{Salvatier {et~al.}(2016)Salvatier, Wiecki, \&
  Fonnesbeck}]{Salvatier2016}
Salvatier, J., Wiecki, T.~V., \& Fonnesbeck, C. 2016, PeerJ Computer Science,
  2, e55

\bibitem[{{Sandage} \& {Tammann}(1987)}]{Sandage1987}
{Sandage}, A. \& {Tammann}, G.~A. 1987, {A Revised Shapley-Ames Catalog of
  Bright Galaxies}

\bibitem[{{Schneider} {et~al.}(2010){Schneider}, {Richards}, {Hall}, {Strauss},
  {Anderson}, {Boroson}, {Ross}, {Shen}, {Brandt}, {Fan}, {Inada}, {Jester},
  {Knapp}, {Krawczyk}, {Thakar}, {Vanden Berk}, {Voges}, {Yanny}, {York},
  {Bahcall}, {Bizyaev}, {Blanton}, {Brewington}, {Brinkmann}, {Eisenstein},
  {Frieman}, {Fukugita}, {Gray}, {Gunn}, {Hibon}, {Ivezi{\'c}}, {Kent}, {Kron},
  {Lee}, {Lupton}, {Malanushenko}, {Malanushenko}, {Oravetz}, {Pan}, {Pier},
  {Price}, {Saxe}, {Schlegel}, {Simmons}, {Snedden}, {SubbaRao}, {Szalay}, \&
  {Weinberg}}]{Schneider2010}
{Schneider}, D.~P., {Richards}, G.~T., {Hall}, P.~B., {et~al.} 2010, \aj, 139,
  2360

\bibitem[{{Schneider} {et~al.}(2015){Schneider}, {Bianchi}, {Valiante},
  {Risaliti}, \& {Salvadori}}]{Schneider2015}
{Schneider}, R., {Bianchi}, S., {Valiante}, R., {Risaliti}, G., \& {Salvadori},
  S. 2015, \aap, 579, A60

\bibitem[{{Scoville} {et~al.}(2007){Scoville}, {Aussel}, {Brusa}, {Capak},
  {Carollo}, {Elvis}, {Giavalisco}, {Guzzo}, {Hasinger}, {Impey}, {Kneib},
  {LeFevre}, {Lilly}, {Mobasher}, {Renzini}, {Rich}, {Sanders}, {Schinnerer},
  {Schminovich}, {Shopbell}, {Taniguchi}, \& {Tyson}}]{Scoville2007}
{Scoville}, N., {Aussel}, H., {Brusa}, M., {et~al.} 2007, \apjs, 172, 1

\bibitem[{{Scoville} {et~al.}(2014){Scoville}, {Aussel}, {Sheth}, {Scott},
  {Sanders}, {Ivison}, {Pope}, {Capak}, {Vanden Bout}, {Manohar}, {Kartaltepe},
  {Robertson}, \& {Lilly}}]{Scoville2014}
{Scoville}, N., {Aussel}, H., {Sheth}, K., {et~al.} 2014, \apj, 783, 84

\bibitem[{{Shakura} \& {Sunyaev}(1973)}]{Shakura1973}
{Shakura}, N.~I. \& {Sunyaev}, R.~A. 1973, \aap, 24, 337

\bibitem[{{Shankar} {et~al.}(2016){Shankar}, {Calderone}, {Knigge}, {Matthews},
  {Buckland}, {Hryniewicz}, {Sivakoff}, {Dai}, {Richardson}, {Riley}, {Gray},
  {La Franca}, {Altamirano}, {Croston}, {Gandhi}, {H{\"o}nig}, {McHardy}, \&
  {Middleton}}]{Shankar2016}
{Shankar}, F., {Calderone}, G., {Knigge}, C., {et~al.} 2016, \apjl, 818, L1

\bibitem[{{Shen} \& {Kelly}(2010)}]{Shen2010}
{Shen}, Y. \& {Kelly}, B.~C. 2010, \apj, 713, 41

\bibitem[{{Shen} {et~al.}(2011){Shen}, {Richards}, {Strauss}, {Hall},
  {Schneider}, {Snedden}, {Bizyaev}, {Brewington}, {Malanushenko},
  {Malanushenko}, {Oravetz}, {Pan}, \& {Simmons}}]{Shen2011}
{Shen}, Y., {Richards}, G.~T., {Strauss}, M.~A., {et~al.} 2011, \apjs, 194, 45

\bibitem[{{Skrutskie} {et~al.}(2006){Skrutskie}, {Cutri}, {Stiening},
  {Weinberg}, {Schneider}, {Carpenter}, {Beichman}, {Capps}, {Chester},
  {Elias}, {Huchra}, {Liebert}, {Lonsdale}, {Monet}, {Price}, {Seitzer},
  {Jarrett}, {Kirkpatrick}, {Gizis}, {Howard}, {Evans}, {Fowler}, {Fullmer},
  {Hurt}, {Light}, {Kopan}, {Marsh}, {McCallon}, {Tam}, {Van Dyk}, \&
  {Wheelock}}]{Skrutskie2006}
{Skrutskie}, M.~F., {Cutri}, R.~M., {Stiening}, R., {et~al.} 2006, \aj, 131,
  1163

\bibitem[{{Stalevski} {et~al.}(2016){Stalevski}, {Ricci}, {Ueda}, {Lira},
  {Fritz}, \& {Baes}}]{Stalevski2016}
{Stalevski}, M., {Ricci}, C., {Ueda}, Y., {et~al.} 2016, \mnras, 458, 2288

\bibitem[{{Talia} {et~al.}(2015){Talia}, {Cimatti}, {Pozzetti}, {Rodighiero},
  {Gruppioni}, {Pozzi}, {Daddi}, {Maraston}, {Mignoli}, \& {Kurk}}]{Talia2015}
{Talia}, M., {Cimatti}, A., {Pozzetti}, L., {et~al.} 2015, \aap, 582, A80

\bibitem[{{Trakhtenbrot} {et~al.}(2017){Trakhtenbrot}, {Ricci}, {Koss},
  {Schawinski}, {Mushotzky}, {Ueda}, {Veilleux}, {Lamperti}, {Oh}, {Treister},
  {Stern}, {Harrison}, {Balokovi{\'c}}, \& {Gehrels}}]{Trakhtenbrot2017}
{Trakhtenbrot}, B., {Ricci}, C., {Koss}, M.~J., {et~al.} 2017, \mnras, 470, 800

\bibitem[{{Treister} {et~al.}(2009){Treister}, {Urry}, \&
  {Virani}}]{Treister2009}
{Treister}, E., {Urry}, C.~M., \& {Virani}, S. 2009, \apj, 696, 110

\bibitem[{{Ueda} {et~al.}(2003){Ueda}, {Akiyama}, {Ohta}, \&
  {Miyaji}}]{Ueda2003}
{Ueda}, Y., {Akiyama}, M., {Ohta}, K., \& {Miyaji}, T. 2003, \apj, 598, 886

\bibitem[{{Urry} \& {Padovani}(1995)}]{Urry1995}
{Urry}, C.~M. \& {Padovani}, P. 1995, \pasp, 107, 803

\bibitem[{{Vanden Berk} {et~al.}(2001){Vanden Berk}, {Richards}, {Bauer},
  {Strauss}, {Schneider}, {Heckman}, {York}, {Hall}, {Fan}, {Knapp},
  {Anderson}, {Annis}, {Bahcall}, {Bernardi}, {Briggs}, {Brinkmann}, {Brunner},
  {Burles}, {Carey}, {Castander}, {Connolly}, {Crocker}, {Csabai}, {Doi},
  {Finkbeiner}, {Friedman}, {Frieman}, {Fukugita}, {Gunn}, {Hennessy},
  {Ivezi{\'c}}, {Kent}, {Kunszt}, {Lamb}, {Leger}, {Long}, {Loveday}, {Lupton},
  {Meiksin}, {Merelli}, {Munn}, {Newberg}, {Newcomb}, {Nichol}, {Owen}, {Pier},
  {Pope}, {Rockosi}, {Schlegel}, {Siegmund}, {Smee}, {Snir}, {Stoughton},
  {Stubbs}, {SubbaRao}, {Szalay}, {Szokoly}, {Tremonti}, {Uomoto}, {Waddell},
  {Yanny}, \& {Zheng}}]{Vandenberk2001}
{Vanden Berk}, D.~E., {Richards}, G.~T., {Bauer}, A., {et~al.} 2001, \aj, 122,
  549

\bibitem[{{Vestergaard} \& {Osmer}(2009)}]{Vestergaard2009}
{Vestergaard}, M. \& {Osmer}, P.~S. 2009, \apj, 699, 800

\bibitem[{{Vestergaard} \& {Peterson}(2006)}]{Vestergaard2006}
{Vestergaard}, M. \& {Peterson}, B.~M. 2006, \apj, 641, 689

\bibitem[{{Viero} {et~al.}(2014){Viero}, {Asboth}, {Roseboom}, {Moncelsi},
  {Marsden}, {Mentuch Cooper}, {Zemcov}, {Addison}, {Baker}, {Beelen}, {Bock},
  {Bridge}, {Conley}, {Devlin}, {Dor{\'e}}, {Farrah}, {Finkelstein},
  {Font-Ribera}, {Geach}, {Gebhardt}, {Gill}, {Glenn}, {Hajian}, {Halpern},
  {Jogee}, {Kurczynski}, {Lapi}, {Negrello}, {Oliver}, {Papovich}, {Quadri},
  {Ross}, {Scott}, {Schulz}, {Somerville}, {Spergel}, {Vieira}, {Wang}, \&
  {Wechsler}}]{Viero2014}
{Viero}, M.~P., {Asboth}, V., {Roseboom}, I.~G., {et~al.} 2014, \apjs, 210, 22

\bibitem[{{Volonteri} {et~al.}(2017){Volonteri}, {Reines}, {Atek}, {Stark}, \&
  {Trebitsch}}]{Volonteri2017}
{Volonteri}, M., {Reines}, A.~E., {Atek}, H., {Stark}, D.~P., \& {Trebitsch},
  M. 2017, \apj, 849, 155

\bibitem[{{Waters} {et~al.}(2016){Waters}, {Wilkins}, {Di Matteo}, {Feng},
  {Croft}, \& {Nagai}}]{Waters2016}
{Waters}, D., {Wilkins}, S.~M., {Di Matteo}, T., {et~al.} 2016, \mnras, 461,
  L51

\bibitem[{{Weingartner} \& {Draine}(2001)}]{Weingartner2001}
{Weingartner}, J.~C. \& {Draine}, B.~T. 2001, \apj, 548, 296

\bibitem[{{Wilkins} {et~al.}(2013){Wilkins}, {Bunker}, {Coulton}, {Croft},
  {Matteo}, {Khandai}, \& {Feng}}]{Wilkins2013}
{Wilkins}, S.~M., {Bunker}, A., {Coulton}, W., {et~al.} 2013, \mnras, 430, 2885

\bibitem[{{Willingale} {et~al.}(2013){Willingale}, {Starling}, {Beardmore},
  {Tanvir}, \& {O'Brien}}]{Willingale2013}
{Willingale}, R., {Starling}, R.~L.~C., {Beardmore}, A.~P., {Tanvir}, N.~R., \&
  {O'Brien}, P.~T. 2013, \mnras, 431, 394

\bibitem[{{Wright} {et~al.}(2010){Wright}, {Eisenhardt}, {Mainzer}, {Ressler},
  {Cutri}, {Jarrett}, {Kirkpatrick}, {Padgett}, {McMillan}, {Skrutskie},
  {Stanford}, {Cohen}, {Walker}, {Mather}, {Leisawitz}, {Gautier}, {McLean},
  {Benford}, {Lonsdale}, {Blain}, {Mendez}, {Irace}, {Duval}, {Liu}, {Royer},
  {Heinrichsen}, {Howard}, {Shannon}, {Kendall}, {Walsh}, {Larsen}, {Cardon},
  {Schick}, {Schwalm}, {Abid}, {Fabinsky}, {Naes}, \& {Tsai}}]{Wright2010}
{Wright}, E.~L., {Eisenhardt}, P.~R.~M., {Mainzer}, A.~K., {et~al.} 2010, \aj,
  140, 1868

\bibitem[{{Zdziarski}(1985)}]{Zdziarski1985}
{Zdziarski}, A.~A. 1985, \apj, 289, 514

\bibitem[{{Zel'dovich} \& {Novikov}(1965)}]{Zeldovich1965}
{Zel'dovich}, Y.~B. \& {Novikov}, I.~D. 1965, Soviet Physics Doklady, 9, 834

\end{thebibliography}

\begin{appendix}
\section{\small{CSTACK}\large{ setup}}\label{sec:stack}
To estimate the amount of the residual X-ray emission from the zombie galaxies by \citet{Capak2015} described in Sect. \ref{sec:chandrastack}, we adopt the following {\scriptsize CSTACK} setup:
\begin{enumerate}
\item use of the {\itshape C}-COSMOS public data set;
\item standard {\itshape Chandra} $0.5 - 2$ and $2 - 8$ keV energy bands;
\item maximum off-axis angle of $15'$;
\item size of $100'' \times 100''$ for the stacked image;
\item radius of $1''.5$ for the source region (corresponding to three times the {\itshape Chandra} angular resolution);
\item inner background radius of $2''$;
\item union of all energy bands for the exclusion of resolved sources;
\item resolved sources excluded from both the source and the background region.
\end{enumerate}
Points 1 and 2 are {\scriptsize CSTACK} defaults. Point 3 ensures that the whole $C$-COSMOS field is explored, avoiding  for example  the exclusion of regions lying at the field-of-view border. Point 4 only generates images large enough to visually identify possible signal excesses concentrated at the image center. Points 5 to 8 allow {\scriptsize CSTACK} to automatically run on source positions and background regions that do not contain any resolved object in the {\itshape C}-COSMOS field: for instance, the weak quasar HZ5 is excluded from the stacking, but the nearby source HZ5a is included due to an angular separation of $1''.8$.
\end{appendix}

\end{document}